\documentclass[]{bytedance_seed}

\usepackage[toc,page,header]{appendix}

\usepackage[utf8]{inputenc} 
\usepackage[T1]{fontenc}    
\usepackage{hyperref}       
\usepackage{url}            
\usepackage{booktabs}       
\usepackage{amsfonts}       
\usepackage{nicefrac}       
\usepackage{microtype}      
\usepackage{xcolor}

\usepackage{amsmath,amssymb}
\usepackage{graphicx}
\usepackage{booktabs}
\usepackage{hyperref}
\usepackage{enumitem}
\usepackage{caption}
\usepackage{arydshln}

\usepackage{algorithm}
\usepackage{algpseudocode}

\newcommand{\ours}{\text{MLPO}}
\newcommand{\agentpolicy}[1]{\pi^{\mathrm{a}_{#1}}}
\newcommand{\allagents}{\boldsymbol{\pi}^{\mathbf{a}}}
\newcommand{\leaderpolicy}{\pi_{\theta}^\mathrm{L}}
\newcommand{\oldleaderpolicy}{\pi^\mathrm{L}_{\theta_{\text{old}}}}
\newcommand{\leaderresponse}{z_L}
\newcommand{\agentresponse}[1]{s_{#1}}
\newcommand{\agent}{\mathrm{a}}

\usepackage[most]{tcolorbox}
\tcbset{
  enhanced,
  colback=gray!5,
  colframe=gray!80,
  fonttitle=\bfseries,
  coltitle=black,
  sharp corners,
  boxrule=0.5pt,
  breakable,
  left=5pt,
  right=5pt,
  top=5pt,
  bottom=5pt,
}

\title{How to Train a Leader: Hierarchical Reasoning in Multi-Agent LLMs}

\author[*,1]{Andrew Estornell}
\author[*,1]{Jean-Francois Ton}
\author[*,1]{Muhammad Faaiz Taufiq}
\author[1]{Hang Li}

\contribution[*]{All authors contributed equally; ordering chosen at random.}
\affiliation[1]{ByteDance Seed}

\abstract{
Large Language Models (LLMs) have achieved strong performance on a wide range of complex reasoning tasks, yet further gains are often possible by leveraging the complementary strengths of multiple models. 
While multi-agent frameworks can improve solution quality by leveraging multiple LLMs, existing methods are often computationally expensive, both at training and inference time. 
In this work, we introduce a hierarchical multi-agent framework that addresses these challenges by training only a single leader LLM to coordinate a team of untrained peer agents. 
To this end, we propose
\textbf{M}ulti-agent guided \textbf{L}eader \textbf{P}olicy \textbf{O}ptimization (MLPO), a novel approach which trains the leader to evaluate and synthesize agent responses without auxiliary value networks or explicit agent feedback.
Leaders trained with MLPO exhibit improved performance not only when interacting with the agent team at inference time, but also enjoy improved performance when deployed in single-agent settings without the team.
Empirical results on Big-Bench Hard (BBH), MATH, and MMLU demonstrate that our framework achieves substantial performance improvements over both single-agent and multi-agent baselines. Our results highlight the effectiveness and efficiency of training a single, flexible leader for collaborative reasoning in multi-agent LLM systems.
}

\correspondence{\email{\{andrew.estornell, jeanfrancois, faaiz.taufiq, lihang.lh\}@bytedance.com}
}

\begin{document}

\maketitle

\section{Introduction}

Large Language Models (LLMs) have rapidly become foundational in natural language processing, demonstrating remarkable versatility across tasks ranging from translation to complex question answering~\citep{brown2020gpt3,bommasani2021foundation}. 
While these models have shown impressive performance in a wide range of tasks, they still struggle with both factual accuracy and complex reasoning 
~\citep{huang2023hallucination,mondorf2024reasoning}.

To address some of these limitations, Multi-Agent approaches in particular have shown promise: by having two or more LLMs collaboratively solve a problem, the group of models may collectively produce better answers than a single model alone~\citep{du2023improving,chen2023reconcile,wu2023autogen,eo2025down,motwani2025malt,jin2025two,estornell2024acccollab,li2024improving,liu2024groupdebate,khan2024debatingpersuasivellmsleads,estornell2024multi,chang2024socrasynth,feng2025heterogeneous,feng2025one,feng2024don,chen2025symbolic,wang2024mixture,li2024smoa,feng2024modular,chan2023chateval}. 
However, current multi-LLM frameworks mostly rely on off-the-shelf models that have not been explicitly trained to collaborate, treating effective collaboration as an emergent property of large models~\citep{du2023improving,chen2023reconcile,yang2025multi,khan2024debatingpersuasivellmsleads}. 
These works assume that general-purpose LLMs are inherently capable of debating, verifying, or correcting each other, and focus on designing (or training) mechanisms to better elicit this behaviour from off-the-shelf models.

In contrast, some recent work has begun to explore training schemes that explicitly tune models for multi-agent collaboration~\citep{acccollab,zhou2025sweet,li2025two,wan2025rema,motwani2025malt,qiu2024towards,park2025maporl}. 
For example, ACC-Collab~\citep{acccollab} jointly trains a two-model team (one actor and one critic) to solve tasks through iterative dialogue. 
Other works have extended this paradigm to teams of two or three similarly specialised models~\citep{zhou2025sweet,li2025two,wan2025rema,motwani2025malt}, demonstrating that training specialised teams can yield performance improvements.
Some approaches further scale to larger groups of trained agents~\citep{qiu2024towards,park2025maporl}. 
However, a key drawback of each of these methods is that they require jointly optimising multiple LLMs, an approach that is computationally expensive and difficult to scale. This raises a fundamental question:

\begin{center}
  \emph{Can we enhance a multi-agent team’s collaborative reasoning by training \textbf{only a single} model, rather than training multiple models?}
\end{center}
In this paper, we answer this question by proposing a novel hierarchical multi-agent architecture for collaborative reasoning, \textbf{M}ulti-agent guided \textbf{L}eader \textbf{P}olicy \textbf{O}ptimization (\ours). 
Our approach draws inspiration from hierarchical structures in multi-agent systems, where a designated leader coordinates a team of subordinate agents~\citep{hong2024metagpt,liu2024hmaw,singh2024malmm,jin2025two,li2025drafts}, and from recent advancements in reinforcement-learning-based reasoning techniques~\citep{grpo}.
Specifically, we introduce a framework in which a single \emph{leader} LLM is trained to solve tasks with the assistance of a team of untrained (off-the-shelf) LLM agents that provide candidate solutions. 
Only the leader is trained, while the supporting team members remain fixed. 
During inference, the leader queries the agent team, aggregates their outputs, and synthesises a final answer.

\ours~can be regarded as multi-agent guided training. During this process, the leader model both strengthens its collaboration with the agent team and broadens its exploration of the solution space by synthesising the diverse candidate answers supplied by the agent team into its own answer.
In RL-based training, providing the leader with a richer pool of candidate solutions broadens its search space, enabling the discovery of more effective reasoning strategies and yielding higher-quality final outputs.

Since our approach involves training only a single model, our multi-agent guided training scheme offers clear advantages, including improved efficiency during training and greater flexibility at inference time. Unlike previous multi-agent methods, which require all agents to participate at test time, our trained leader can effectively function independently. 
We empirically observe that our training pipeline not only enhances the leader's performance when interacting with the agent team at inference time, but also enhances its performance as an individual model operating without the team.
\emph{In other words, our training approach enhances both collaborative and zero-shot performance.} 
Consequently, when inference-time cost is a constraint, the leader alone can be deployed and still achieves competitive results, although full collaboration generally provides the best performance

We evaluate our hierarchical leader team architecture using \emph{7--9B}-parameter models on three challenging reasoning suites: \textsc{Big-Bench Hard}~(BBH) \cite{bbh}, \textsc{MATH} \cite{mathdataset}, and \textsc{MMLU} \cite{mmlu}.
Against two classes of baselines (i) strong \emph{prompt-only} methods such as Chain-of-Thought and multi-agent debate with off-the-shelf LLMs~\cite{cot,du2023improving}, and (ii) \emph{trained} variants that employ the same agents but use different training techniques as well as components \cite{acccollab,maurya2024selectllm,kumar2024training, grpo}. 

The contributions of this paper are as follows:
\begin{itemize}
    \item \textbf{Hierarchical framework}: We propose a hierarchical multi-agent architecture where a single leader LLM is explicitly trained to coordinate a team of untrained peer LLMs for collaborative reasoning.
    \item \textbf{Multi-agent guided GRPO objective}: We introduce a novel GRPO-based approach, which we coin \ours~, enabling the leader to implicitly evaluate and synthesize agent responses without the need for auxiliary value networks or explicit agent feedback signals.
    \item \textbf{Strong empirical results}: Extensive experiments demonstrate that our hierarchical multi-agent approach significantly outperforms existing baselines, including both trained and untrained single-agent and multi-agent methods.
    \item \textbf{Systematic ablation study}: We perform thorough ablations on team composition, aggregation methods, and sampling strategies, providing insights into their individual contributions and guiding further improvements to our framework.
\end{itemize}

\section{Related Work}
Our work is mostly closely related to the areas of multi-LLM collaboration and iterative improvement. 

\paragraph{\textbf{Multi-LLM Collaboration}} Similar our proposed method, several other works have investigated the use of leveraging multiple LLMs to improve answer quality. 
These works can be loosely grouped into two categories: \emph{collaborative methods}, which engage multiple LLMs in collaboration, often in the form of iterative discussion \cite{du2023improving,estornell2024acccollab,li2024improving,liu2024groupdebate,liang2023encouraging,khan2024debatingpersuasivellmsleads,estornell2024multi,chang2024socrasynth,feng2025heterogeneous,feng2025one,feng2024don,li2023prd,ye2025x,yang2025agentnet}, and \emph{aggregation methods} which attempt to bootstrap multiple LLMs (responses) together without direct collaboration \cite{dai2024cost,yang2025multi,jiang2023llm,huang2024ensemble}. 
There are some methods that fall into both categories, such as mixture of agents (MoA) \cite{chen2025symbolic,wang2024mixture,li2024smoa}.

Several unique paradigms have been proposed within the space of collaborative multi-LLM approaches. 
Notably, \cite{du2023improving} proposes a system of multi-agent debate in which a team of LLMs directly communicates with one another over rounds of discussion. 
Other works have extended this basic framework with a focus on designing mechanisms to enhance the collaborative ability of off-the-shelf-models; \cite{khan2024debatingpersuasivellmsleads} introduces a judge/moderator LLM, \cite{chen2023reconcile,zhang2024chain} propose alternating the order of model interaction, \cite{li2023prd} allows LLMs to score each other's responses in between interaction rounds, many other such mechanisms have been proposed such as \cite{liang2023encouraging,feng2025heterogeneous,feng2025one,feng2024don,li2023prd,ye2025x,yang2025agentnet,liu2024hmaw}. 
While effective in certain settings, prior approaches do not directly train models to collaborate. 
Instead, they aim to improve the collaboration of off-the-shelf models through either structured interaction schemes (e.g., two LLMs debating with a third acting as a judge) or learned selection mechanisms (e.g., a classifier that selects a single response from a pool of candidate outputs).
Diverging from this work, our proposal involves training a reasoning-based leader LLM which critically analyzes the team responses, thereby explicitly guiding the team towards the right direction when needed.

More recently, several recent works have proposed training schemes to directly enhance the collaborative ability of LLMs. 
In particular, ACC-Collab \cite{acccollab} designs a scheme for training a two model team consisting of an actor and critic agent. Building upon ACC-Collab, other works have also designed two-agent  \cite{zhou2025sweet,li2025two,wan2025rema,ma2024coevolving} (and three agent \cite{motwani2025malt}) training paradigms.
Extending this further, other works such as  \cite{qiu2024towards,park2025maporl,liao2025marft} train multiple models to collaborate to directly collaborate.
However, these training schemes can be exceedingly expensive as they require training multiple models simultaneously or alternatively. 
Our framework aims to make multi-agent training more efficient by training only a \textit{single leader agent} without the requirement of training the other models in the team.

\paragraph{\textbf{Self-Improvement}} 
Our work is also related to self-improvement. 
Techniques in this area aim to improve LLM efficacy by iteratively refining given answers. 
This refinement can come in many forms, from prompting models to improve their previous answers \cite{madaan2023self}, to iterative exploration or expansion of their reasoning \cite{cot,yao2023tree,besta2024graph}, and even to directly training advanced single-shot reasoning capabilities into the model itself \cite{grpo,shao2024deepseekmath}.
Of particular relevance to our work is that of \cite{kumar2024training} which trains a model to directly update and refine its own responses. This can be thought of as an explicit form of the type of \textit{backtracking} behaviour found in modern reasoning models \cite{jaech2024openai,grpo,qin2024o1,yang2025qwen3}. 
In our method, the leader agent learns to perform a type of macro-update to the agent team responses, while using those responses to synthesize its own refined answer and feedback.

\section{Methodology}

\subsection{Hierarchical Multi-Agent Inference Pipeline}

\begin{figure}[]
    \centering
    \includegraphics[width=0.5\textwidth]{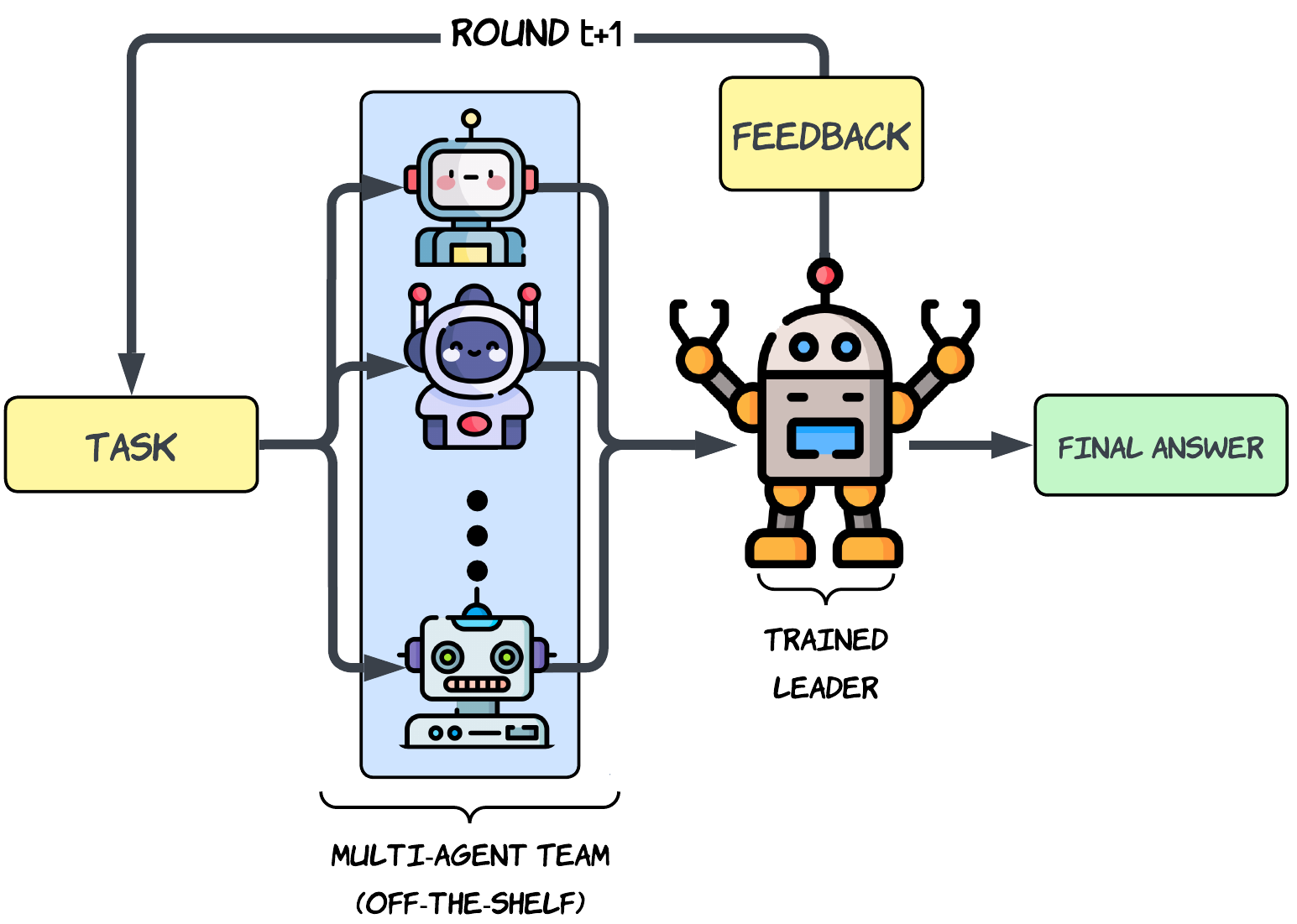}
    \caption{Overview of the proposed hierarchical multi-agent inference architecture. A user prompt is first processed by a team of $K$ off-the-shelf agents whose intermediate generations are forwarded to a leader model trained using our \ours~pipeline. The leader’s output is then returned to the agent team together with the initial prompt, and this exchange is repeated for $T$ iterations before the final answer is produced.}
    \label{fig:inference-pipeline}
\end{figure}
\paragraph{\textbf{Multi-Agent Team Setup}} Our methodology utilizes a hierarchical multi-agent architecture comprising a single leader model $L$ and a team of $K$ off-the-shelf agent models, denoted as $\{\agent_1, \agent_2, \ldots, \agent_K\}$. Although we set $K=3$ in our experiments, this framework generalizes 
to any number of agents. 
Notably, we show that training only the leader model, without requiring that the agent team also be trained, is sufficient to achieve strong collaborative effectiveness and also makes our method considerably more scalable and practical than approaches requiring all agents to be trained.

\paragraph{\textbf{Multi-Agent Team Inference}} 
The inference process unfolds iteratively over $T$ rounds, as summarised in Figure \ref{fig:inference-pipeline}. In the initial round (round 0), each agent independently generates a preliminary solution to a given task input $x$. We denote Agent $i$'s policy by $\agentpolicy{i}$ and its $0$'th round preliminary solution by $\agentresponse{i}^{(0)} \sim \agentpolicy{i}(x)$. The leader model, denoted by $\leaderpolicy$, then synthesizes these initial agent responses into its structured output, denoted by $\leaderresponse^{(0)}$, i.e.,
\[
\leaderresponse^{(0)} \sim \leaderpolicy\left(x, \agentresponse{1}^{(0)}, \agentresponse{2}^{(0)}, \ldots, \agentresponse{K}^{(0)}\right).
\] 
This output encapsulates detailed reasoning within specific \texttt{<think>} tags and presents a consolidated solution within \texttt{<answer>} tags.

In subsequent rounds ($t = 1, 2, \ldots, T-1$), each agent revises its solution based on the leader's previous output and its own earlier response:
\[
\agentresponse{i}^{(t)} \sim \agentpolicy{i}\left(x, \agentresponse{i}^{(t-1)}, \leaderresponse^{(t-1)}\right).
\]
The leader then integrates these updated agent solutions to refine its reasoning and produce an updated consolidated output:
\[
\leaderresponse^{(t)} \sim \leaderpolicy\left(x, \agentresponse{1}^{(t)}, \agentresponse{2}^{(t)}, \ldots, \agentresponse{K}^{(t)}\right).
\]
This iterative cycle of feedback and refinement continues until the final round, after which the final solution is extracted from the leader’s output.

\subsection{Multi-Agent Guided Leader Training}\label{sec:multi_agent_training}

Our training approach comprises two phases similar to \cite{guo2025deepseek,team2025kimi}: Supervised Fine-Tuning (SFT), designed to develop the leader's natural backtracking and self-correction abilities, followed by Group Relative Policy Optimization (GRPO) \cite{grpo}, aimed at enhancing multi-agent collaboration skills. Notably, our experimental results using Qwen2.5 7B Instruct as the leader indicate that the SFT step, while beneficial, is not strictly necessary before GRPO. Specifically, we observe that the largest accuracy gains for the leader model arise primarily from the GRPO phase itself, whereas preceding GRPO with SFT provides only modest incremental improvements (typically 1-2\% accuracy increase).

\subsubsection{Supervised Fine-Tuning} 
The SFT phase aims to train the leader model to reinforce its aggregation, self-correction and backtracking behaviours, which are crucial skills for effective collaboration and error recovery. To this end, we first run our multi-agent system with an untrained leader model. For each task in the training data, we generate one round of agent responses. Next, we query the untrained leader to use these agent responses to synthesize a final solution. For each set of agent responses, we query the leader $16$ times to obtain diverse solutions to each prompt. We then filter out the tasks where all $16$ leader responses are incorrect, as these likely represent tasks beyond the model's current capabilities.

For the remaining tasks, we construct a new SFT dataset comprising generations with self-correction and backtracking, inspired by previous approaches explored in recent literature \cite{qin2024o1,guo2025deepseek,team2025kimi}. To achieve this, we first randomly select one correct and one incorrect leader response from the $16$ leader generations. Next, an untrained leader model is prompted with these selected responses to generate a ``\textit{backtracked solution}'' that mimics natural reasoning patterns, i.e. beginning with plausible but incorrect reasoning, then naturally self-correcting to arrive at the correct answer. The prompts explicitly encourage natural self-correction phrases such as ``Wait, that doesn't seem right'' or ``Let me reconsider this approach''. For tasks where all 16 responses are correct, we directly include the correct solutions without artificial backtracking. The leader model is then fine-tuned on this dataset of backtracked and correct solutions, aiming to enhance its ability to naturally self-correct and recover from reasoning errors.  

\subsubsection{Multi-Agent Guided Leader Policy Optimization}

\begin{figure}[]
    \centering
\includegraphics[width=0.95\linewidth]{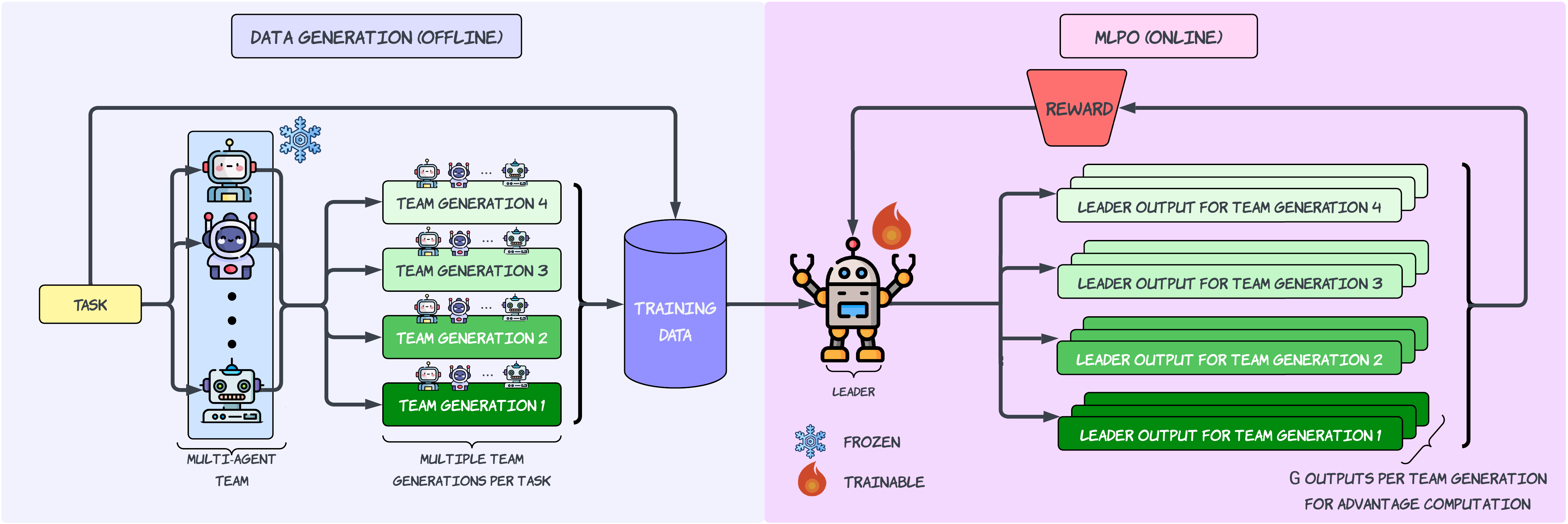}
    \caption{Outline of our Multi-agent guided Leader Policy Optimization (MLPO) pipeline.}
    \label{fig:training_pipeline}
\end{figure}
Following SFT, we employ 
our variant of Group Relative Policy Optimization (GRPO) \cite{grpo}, which we refer to as 
Multi-agent guided Leader Policy Optimization (\ours), 
to train the leader $\leaderpolicy$ specifically for multi-agent collaboration. 
In this phase, the goal is to learn effective aggregation and synthesis strategies. 
In classical GRPO, each prompt simply comprises the task to be solved (along with any accompanying instructions, such as ``let's think step-by-step'', etc.). In contrast, for MLPO, each leader's prompt also includes the agents' solutions to the given task. 
The loss for MLPO can therefore be expressed as:
\begin{align*}
\label{eq:grpo-objective}
\mathcal{J}_{\text{MLPO}}(\theta)
=& \, \mathbb{E}_{x \sim P_{X},\, \mathbf{s} \sim \allagents(x), \, \{o_i\}_{i=1}^{G} \sim \oldleaderpolicy(x, \mathbf{s})} \Biggl[
  \frac{1}{\sum_{i=1}^{G}\lvert o_i \rvert}
  \sum_{i=1}^{G}\sum_{t=1}^{\lvert o_i \rvert}
  \Biggl\{
      \min\Bigl(
      r_{i,t}
        \,
        \hat A_{i,t},
        \;
        \operatorname{clip}\Bigl(
          r_{i,t},\,1-\varepsilon,\,1+\varepsilon
        \Bigr)\,\hat A_{i,t}
      \Bigr)
  \Biggr\}
\Biggr].
\end{align*}
Here, $\textstyle r_{i,t} = \frac{\leaderpolicy\left(o_{i,t}\mid x, \mathbf{s}, o_{i,<t}\right)}{\oldleaderpolicy\left(o_{i,t}\mid x, \mathbf{s}, o_{i,<t}\right)}$ is the importance ratio and $\hat A_{i,t} = R_i - \text{mean}(\textbf{R})$ denotes the advantage of the $i$'th response where $R_i$ is the corresponding reward. 
Moreover, $x$ represents the task input (e.g. a mathematics problem), $\mathbf{s} = \{s^{0}_i\}_{i=1}^K$ denotes the set of agent solutions and $o_i$ is a sample from the current leader $\oldleaderpolicy$. 
The hyperparameter \(\varepsilon\) denotes the PPO-style clipping threshold \cite{schulman2017proximalpolicyoptimizationalgorithms, grpo}.
Note that we have used the modifications to the GRPO loss proposed in Dr. GRPO \cite{liu2025understandingr1zeroliketrainingcritical} due to its improved training efficiency and stability.

\paragraph{\textbf{Training Data Generation}}
To generate our \ours~ training data, we query each of the $K$ agents to independently generate $4$ solutions to each task in the training split (as shown in Figure \ref{fig:training_pipeline}). This results in a total of $4K$ responses per task.
From these responses, we create 4 distinct training prompts per task, each consisting of the original task paired with exactly one response from each agent. This data generation strategy has been validated experimentally in prior work \cite{yao2025r1} as well as in our ablation study in Section \ref{sec:best_prac}, demonstrating that generating multiple training prompts per task significantly improves the test accuracy of the leader model. Moreover, our experiments reveal that increasing the number of responses beyond 4 per agent per task does not substantially further enhance the performance of the leader model.

Next, we also filter out ``easy'' tasks, specifically those tasks where at least 75\% of the $4K$ agent responses are correct. This filtering directs training attention toward scenarios requiring strong leadership and aggregation skills, which further helps boost the leader's performance, as we demonstrate empirically in Section \ref{sec:best_prac}. Additionally, while the tasks are randomly shuffled during training, the four prompts corresponding to each task remain grouped. This strategic ordering allows the leader model to observe different agent team responses (and reasoning) to the same task consecutively, and enhances learning efficiency. Figure \ref{fig:reward} show that this grouping of prompts by tasks makes training much more stable compared to random ordering of data.

Throughout the \ours~training process, the leader model receives prompts containing the original tasks and agent responses. The leader must then produce structured outputs, clearly separating their detailed reasoning within \texttt{<think>} tags and providing definitive answers within \texttt{<answer>} tags. Additionally, similar to various recent works on RL finetuning \cite{grpo, liu2025understandingr1zeroliketrainingcritical, yu2025dapoopensourcellmreinforcement}, we used a simple heuristic reward to avoid any reward hacking, which consisted of a $0$ and $+1$ reward for incorrect and correct answers respectively, as well as a formatting reward for properly structuring responses within the \texttt{<think>} and \texttt{<answer>} tags.

\paragraph{\textbf{Multi-round training with MLPO+}}
Initially, our pipeline trains the leader solely on agent-team responses from round 0, i.e., responses generated before any interaction between the leader and the agent team has occurred. 
This setup enables efficient offline training, as round-0 responses can be precomputed prior to leader optimization. 
While this training scheme is already effective, it does not leverage the full distribution of team responses that arise at inference.

To address this, we also consider a multi-round extension of MLPO, which we call MLPO+, that incorporates an additional training phase using team responses from later rounds ($t > 0$). 
These responses resemble more closely those seen during inference, as they are shaped by interaction with the leader.
To operationalize MLPO+ efficiently, we treat the multi-round training as a continuation phase that follows the initial MLPO training. 
After the leader is trained using round-0 responses, it is deployed to interact with the agent team over multiple rounds. These interactions are then recorded and used to construct a new training set.
Importantly, this additional data can be generated offline, just like the round-0 responses, introducing little extra overhead.

In practice, we first filter out ``easy" tasks (those where the leader and team achieve over 50\% accuracy after two rounds). 
For the remaining tasks, we construct a new dataset consisting of two sets of agent responses from rounds 0, 1, and 2. The leader is then further optimized on this dataset using the same MLPO loss. 
This setup allows the leader to refine its strategy based on more realistic team behaviors that emerge during iterative interaction.

\paragraph{\textbf{Flexible Zero-Shot and Multi-Agent Inference}}
Interestingly, the benefits of our multi-agent guided training extend beyond collaborative settings at inference time. We discovered through our empirical analysis (detailed in Section~\ref{subsec:zeroshot}) that 
training the leader model using our \ours~framework significantly enhances its zero-shot inference capabilities as well, even without employing the agent team during inference. This enables us to achieve higher accuracy than models trained with standard single-agent GRPO approaches, at no additional computational cost during inference. Moreover, when inference-time compute resources permit, deploying the leader alongside the agent team further boosts accuracy, outperforming state-of-the-art multi-agent methods. Thus, our framework flexibly provides improved accuracy in both single- and multi-agent inference settings.

\section{Experimental Results}
\subsection{Experimental Setup}

\textbf{Datasets:} We evaluate our hierarchical multi-agent collaboration framework on three benchmarks that cover a variety of factual- and reasoning-based tasks: Big-Bench Hard (\textbf{BBH}) \cite{bbh},  \textbf{MATH} \cite{mathdataset}, and Massive Multitask Language Understanding (\textbf{MMLU}) \cite{mmlu}.

\textbf{Model Configuration:} In our main experiments, we use Qwen2.5 7B Instruct (\textbf{Qwen-2.5} \cite{qwen2025qwen25technicalreport}) as our leader model, with a heterogeneous agent team comprising Llama 3.1 8B Instruct (\textbf{Llama-3.1} \cite{grattafiori2024llama3herdmodels}), Gemma2 9B Instruct (\textbf{Gemma-2} \cite{gemmateam2024gemma2improvingopen}), and Qwen2.5 7B Instruct. 
In  Section \ref{sec:team_dynamics} we provide results for different choices of agent teams, and in Appendix \ref{sec:sup_diff_leader} we provide results for different leaders (Gemma-2 and Llama-3.1).

\textbf{Baselines:} We compare our method \ours~with several baselines. 
We delineate these baselines into two main categories: training-based and training-free. 
For training-free baselines, we compare to \textbf{Self-Reflection} \cite{madaan2023self}, Multi-Agent Debate (\textbf{MAD}) \cite{du2023improving}, our pipeline with \textbf{untrained leader}, and \textbf{zero-shot} inference, which invokes an untrained single LLM once with no multi-agent team support.
For training-based baselines, we compare to \textbf{ACC-Collab} \cite{acccollab}, \textbf{SelectLLM} \cite{maurya2024selectllm}, \textbf{SCoRe} \cite{kumar2024training}, a \textbf{Deferral leader} trained to always defer to one agent in the team, and \textbf{zero-shot GRPO} \cite{grpo} which invokes a single LLM (trained using standard zero-shot GRPO) once with no multi-agent team support. 
For all iterative methods, we fix the number of inference rounds as $T=5$.

\subsection{Main Results}
In this section, we present an extensive empirical analysis designed to validate and display the capabilities of our proposed \ours~framework. Specifically, our experiments address the following points:

\begin{itemize}
    \item \textbf{Our Proposed \ours~vs. Existing Baselines:} We evaluate our method extensively against existing approaches, demonstrating how our collaborative reasoning approach significantly enhances performance.

    \item \textbf{Test-Time Scaling for Improved Performance:} We investigate how additional computational resources during inference can further amplify the performance gains of our proposed multi-agent team.

    \item \textbf{Multi-agent Guided Training with Zero-Shot Inference}: 
    We demonstrate that our multi-agent guided training pipeline (\ours) also substantially enhances zero-shot performance, surpassing models trained with conventional single-agent GRPO methods without incurring extra inference-time costs.

    \item \textbf{Leader-Agent Interaction Dynamics:} We analyze the interaction dynamics between the leader model and its agent team, showing how our method effectively utilizes agent diversity to achieve robust, accurate, and consistently superior outcomes across various task categories and difficulty levels.
\end{itemize}

\subsubsection{Our Proposed \ours~vs. Existing Baselines}
\begin{table}[htb]
\centering
\renewcommand{\arraystretch}{1.2}
\setlength{\tabcolsep}{8pt}
\begin{footnotesize}
\begin{sc}
\begin{tabular}{l l c c c}
\toprule
Type & Method & MMLU & BBH & MATH \\
\midrule
\multirow{5}{*}{\shortstack[l]{Training-\\free}} 
& Zero-Shot$^*$ & 0.734$_{\pm0.006}$ & 0.733$_{\pm0.010}$ & 0.666$_{\pm0.002}$ \\
& Self-Reflect \cite{madaan2023self} & 0.746$_{\pm0.003}$ & 0.766$_{\pm0.011}$ & 0.681$_{\pm0.003}$ \\
& MAD (Div) \cite{du2023improving} & 0.771$_{\pm0.002}$ & 0.785$_{\pm0.002}$ & 0.653$_{\pm0.009}$ \\
& MAD (Hom) \cite{du2023improving} & 0.759$_{\pm0.003}$ & 0.799$_{\pm0.016}$ & 0.720$_{\pm0.009}$ \\
& Untrained Leader & 0.731$_{\pm0.008}$ & 0.764$_{\pm0.007}$ & 0.697$_{\pm0.007}$ \\
\hline
\multirow{6}{*}{\shortstack[l]{Training-\\based}} 
& ACC-Collab \cite{acccollab} & 0.761$_{\pm0.004}$ & 0.802$_{\pm0.003}$ & 0.698$_{\pm0.008}$ \\
& SelectLLM$^*$ \cite{maurya2024selectllm} & 0.770$_{\pm0.001}$ & 0.768$_{\pm0.001}$ & 0.651$_{\pm0.001}$ \\
& SCoRe \cite{kumar2024training} & 0.752$_{\pm0.005}$ & 0.828$_{\pm0.004}$ & 0.721$_{\pm0.004}$ \\
& Deferral Leader & 0.750$_{\pm0.004}$ & 0.775$_{\pm0.007}$ & 0.720$_{\pm0.003}$ \\
& Zero-Shot GRPO$^*$ \cite{grpo} & 0.742$_{\pm0.004}$ & 0.791$_{\pm0.008}$ & 0.712$_{\pm0.005}$ \\
& Self-Reflect GRPO \cite{grpo} + \cite{madaan2023self} & 0.747$_{\pm0.003}$ & 0.814$_{\pm0.006}$ & 0.718$_{\pm0.006}$ \\
\hline
\multirow{4}{*}{\shortstack[l]{Our\\results}} 
& SFT + \ours~Leader (zero-shot acc w/o team)$^*$ & 0.757$_{\pm0.007}$ & 0.855$_{\pm0.008}$ & 0.729$_{\pm0.008}$ \\
\cdashline{2-5}
& SFT Leader & 0.741$_{\pm0.008}$ & 0.762$_{\pm0.008}$ & 0.713$_{\pm0.007}$ \\
& \ours~Leader  & 0.759$_{\pm0.006}$ & 0.865$_{\pm0.005}$ & 0.750$_{\pm0.009}$ \\
& SFT + \ours~Leader  & \textbf{0.782}$_{\pm0.006}$ & \textbf{0.882}$_{\pm0.005}$ & \textbf{0.762}$_{\pm0.005}$ \\
\bottomrule
\end{tabular}
\end{sc}
\end{footnotesize}
\caption{Accuracy of each method on MMLU, BBH, and MATH benchmarks, with $\pm2$ standard errors. Here, $^*$ denotes methods not using iterative inference. 
}
\label{tab:benchmark-results}
\end{table}
Having outlined our model configuration and selected baselines, we now present the empirical evaluation of our proposed method. Specifically, we systematically investigate our \ours~approach relative to standard single-agent and multi-agent baselines, highlighting the advantages of incorporating collaborative reasoning during both training and inference. Table~\ref{tab:benchmark-results} summarizes our experimental results comparing \ours~with existing baselines.

Our method substantially outperforms both single-agent training baselines (e.g., SCoRe \cite{kumar2024training}, GRPO \cite{grpo}) and multi-agent training baselines (e.g., ACC-Collab \cite{acccollab}, SelectLLM \cite{maurya2024selectllm}, Deferral Leader). 
This performance gain stems from the fact that, during both training and inference, the leader is guided by a diverse set of strategies and solutions proposed by the agent team. 
In contrast, single-agent approaches operate in isolation, lacking access to this breadth of insight, as they must rely on their own solutions. 

Beyond demonstrating superior performance compared to baseline methods, we make two additional key observations regarding the behaviour of our leader. First, upon evaluating the training paradigms, we observe that supervised fine-tuning (SFT) alone does not consistently provide substantial performance improvements over an untrained leader, underscoring the importance of employing reinforcement learning within our MLPO framework. Second, to ensure our leader genuinely develops a strategic decision-making capability rather than merely replicating accurate agent responses, we compare it against two baselines: SelectLLM \cite{kumar2024training}, which identifies the optimal agent subset based solely on the input question, and the Deferral Leader, which selects the best response given both the question and agent outputs. Our SFT + \ours'ed leader consistently outperforms these baselines across all evaluated domains, providing evidence of its capability to formulate independent strategies beyond mere imitation. This conclusion is further corroborated by Figure \ref{fig:leaderCalPerAgent}, where our leader maintains significantly higher accuracy even in scenarios where all individual agents fail.

However, a natural question to ask is whether our improvements are merely a consequence of our method being allowed to use more generations during inference. To address this concern, we perform thorough ablation studies in the following subsections, which show that our method retains its advantage even when controlling for inference-time compute.

\subsubsection{Test-Time Scaling for Improved Performance}
\begin{figure}[tbh]
    \centering
\includegraphics[width=0.95\linewidth]{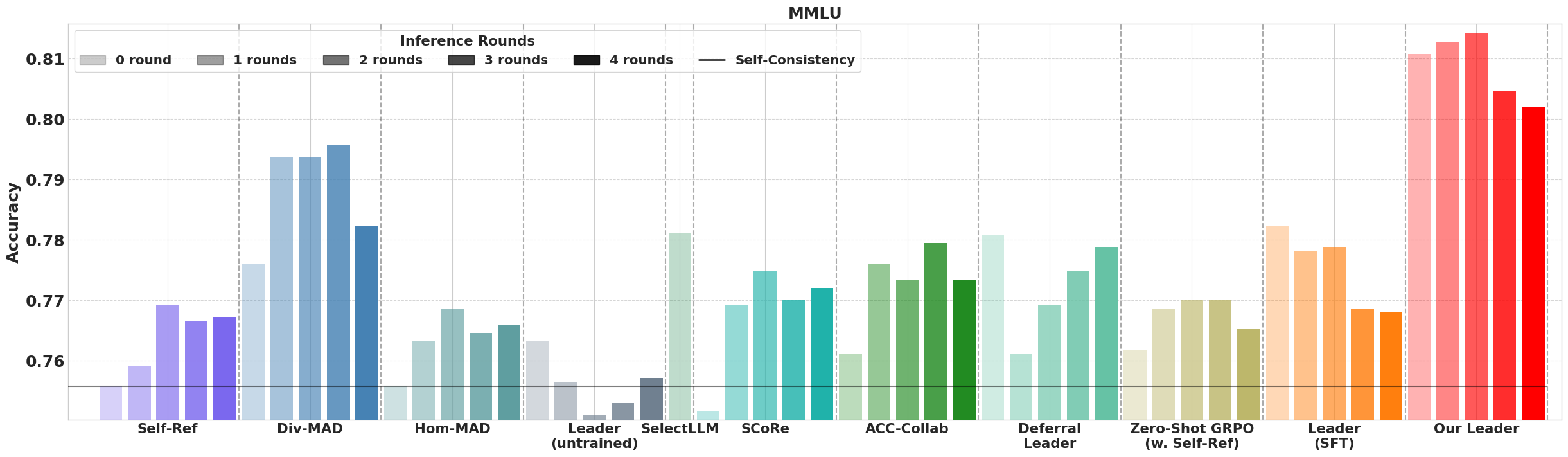}
    \includegraphics[width=0.95\linewidth]{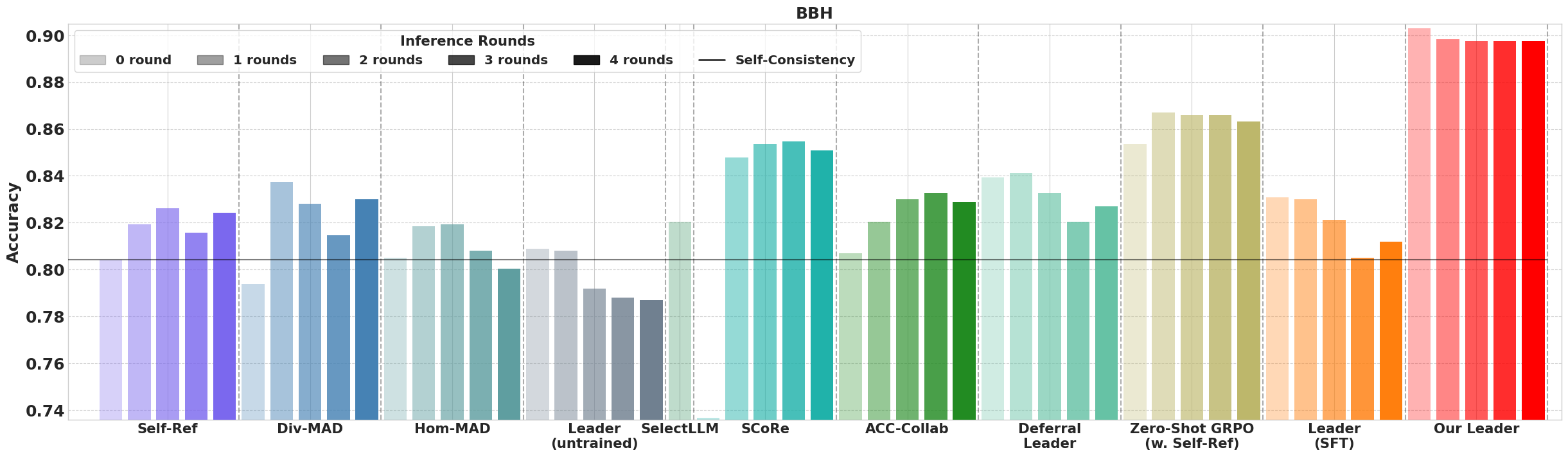}
    \includegraphics[width=0.95\linewidth]{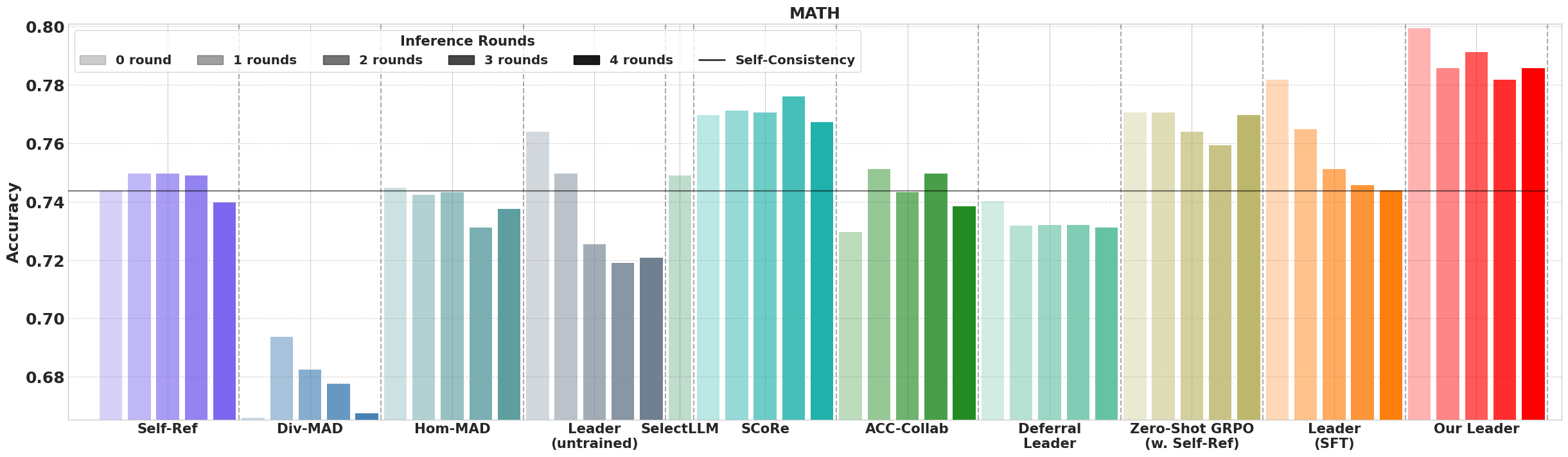}
    \caption{
    Majority vote performance when each method can use at most 40 total LLM generation samples. 
    For iterative methods, we only use the final round answers for majority vote; SelectLLM is not an iterative method. 
    Traditional self-consistency \cite{wang2022self} is shown as horizontal line. 
    The leader model trained with SFT + MLPO is denoted as Our Leader.
    }
    \label{fig:majVote}
\end{figure}

Here, we investigate performance when all methods are allotted the same amount of inference time compute, and explore how additional inference-time compute can be leveraged to further improve the performance of models trained with \ours~under a fixed inference budget.
In particular, prior work has shown that self-consistency, i.e., majority voting over multiple generations, can enhance model accuracy \cite{wang2022self}. For fair comparison, we constrain all methods (including iterative and multi-agent baselines) to a maximum of 40 total LLM generations at inference time. For iterative methods, we apply majority voting only over responses from the final inference round, ensuring that the highest-quality outputs are used. For example, in MAD with 3 agents, we take each agent's final-round response as a vote, resulting in 3 votes for each run of the MAD pipeline, whereas, for our pipeline, we only use the leader’s answer in the final round as a vote for each run.

Figure~\ref{fig:majVote} reports the accuracy of each method under this consistent sample budget with majority voting. 
These results show that our method achieves superior performance across all three datasets when inference-time compute is scaled. 
Notably, our approach requires only two rounds of interaction to reach this performance level, outperforming baselines that require up to five rounds.

We hypothesize that this improvement arises from how different methods trade off bias and variance when scaling inference-time compute. 
In general, iterative pipelines can be scaled either sequentially by increasing the number of rounds per run, or in parallel by generating more independent runs which are then bootstrapped. 
Sequential scaling tends to reduce variance (as agents converge on a shared answer), but may increase bias if the model converges to an incorrect solution \cite{xiong2023examining,estornell2024multi,chang2024socrasynth}. 
In contrast, parallel scaling introduces more response diversity (i.e., higher variance), which can mitigate bias and often improves performance when using techniques such as majority voting \cite{wang2022self,chen2023universal,smit2023should}.

Our method may benefit particularly from parallel scaling due to having two distinct sources of diversity: the variability in responses from the agent team and the stochasticity in the leader's own generations. 
This dual source of response diversity could help explain why our pipeline yields higher majority-vote accuracy under fixed sample budgets, compared to baselines that lack one or both of these mechanisms.

\subsubsection{Enhanced Zero-Shot Capabilities}\label{subsec:zeroshot}
\begin{figure}[t]
    \centering
    \includegraphics[width=0.33\linewidth]{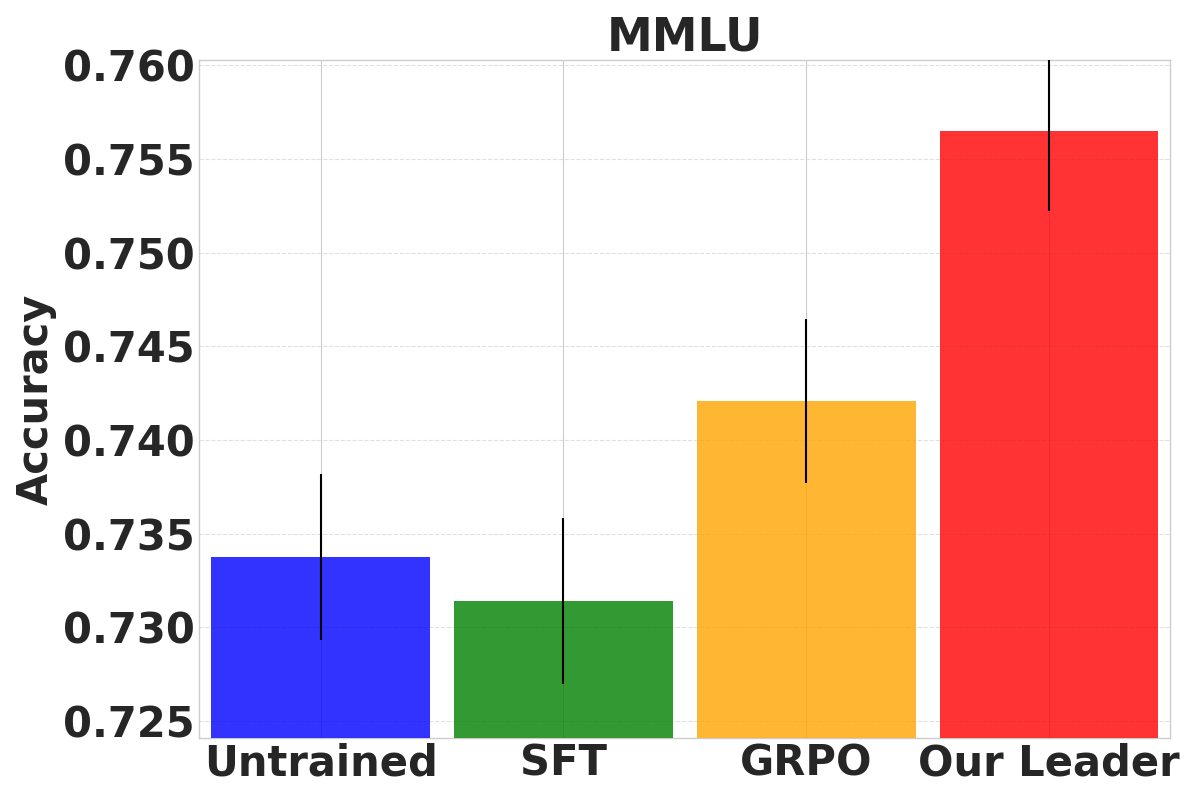}\includegraphics[width=0.33\linewidth]{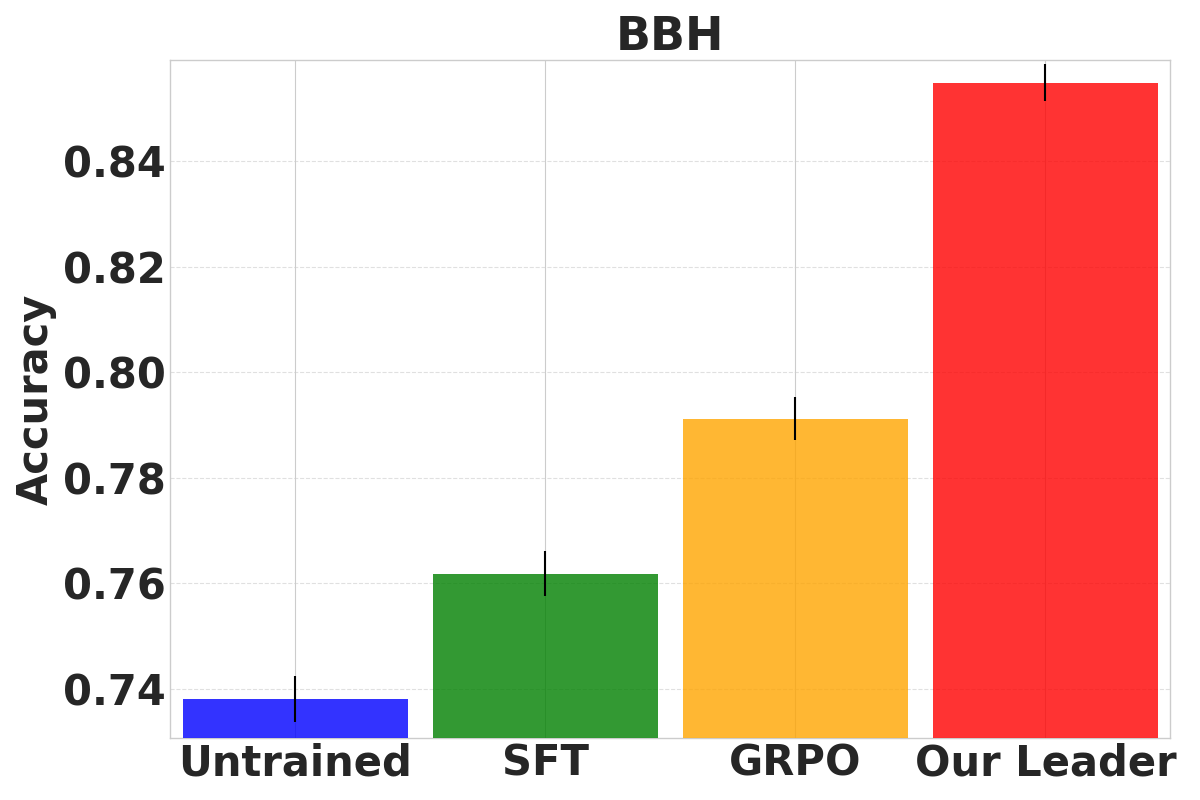}\includegraphics[width=0.33\linewidth]{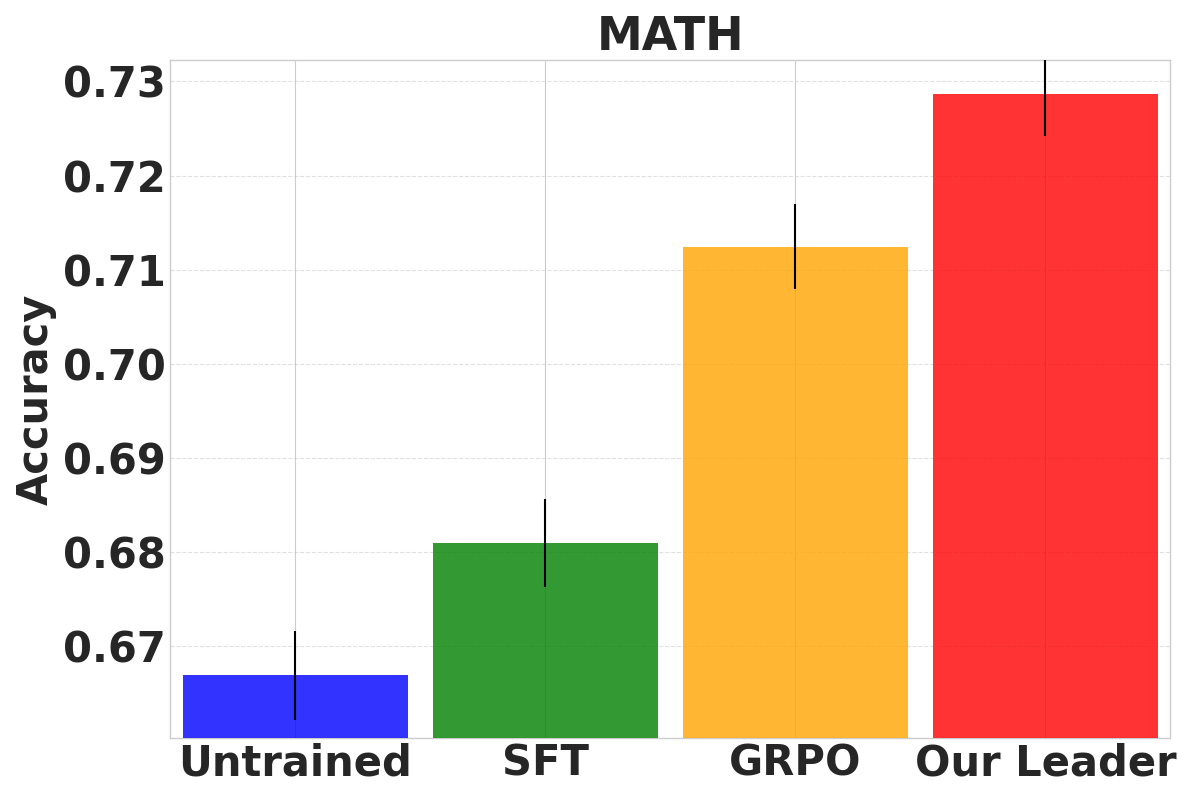}
    \caption{Zeroshot performance of Qwen when untrained, SFTed or GRPOed on zero-shot data, and trained via our MLPO pipeline (Our Leader). The error-bars show $\pm2$ standard errors.}
    \label{fig:zeroShotBar}
\end{figure}

Remarkably, as mentioned previously, our multi-agent trained leader also exhibits superior performance even during zero-shot inference (without multi-agent assistance), outperforming models trained with standard SFT + GRPO pipelines, hence incurring \textbf{no additional inference cost}; see Figure \ref{fig:zeroShotBar}.

This finding suggests that \ours~enhances ``\textit{knowledge acquisition}" during training, potentially through improved exploration of alternative solutions proposed by the agent team. 
The key difference between our multi-agent guided training framework \ours, and standard single-agent training frameworks such as GRPO, lies in the training prompts: the former includes agent solutions alongside the question, while the latter contains only the question and instructions. 
The inclusion of diverse agent responses appears to create a richer training signal that improves the leader's reasoning capabilities even when later deployed in isolation. 

An astute reader may ask if this enhanced ``knowledge acquisition'' is primarily responsible for the improved performance of our multi-agent system. 
In other words, \textit{could it be that even when deployed with the multi-agent team, the trained leader is simply ignoring the agent responses at inference time and using its acquired knowledge to drive the accuracy gains}? 
Our results in Table \ref{tab:benchmark-results} show that this is not the case, since the leader's accuracy improves even further when it is deployed with the multi-agent team at inference time (as compared to its zero-shot inference performance). 
Not only has the leader acquired knowledge, but also utilizes its implicit evaluation and aggregation skills to synthesize a higher-quality final solution when given access to the agent team's responses.

\subsubsection{Leader and Agent Team Interaction Dynamic}\label{sec:team_dynamics}
\begin{figure}[tbh]
    \centering
    \includegraphics[width=0.99\linewidth]{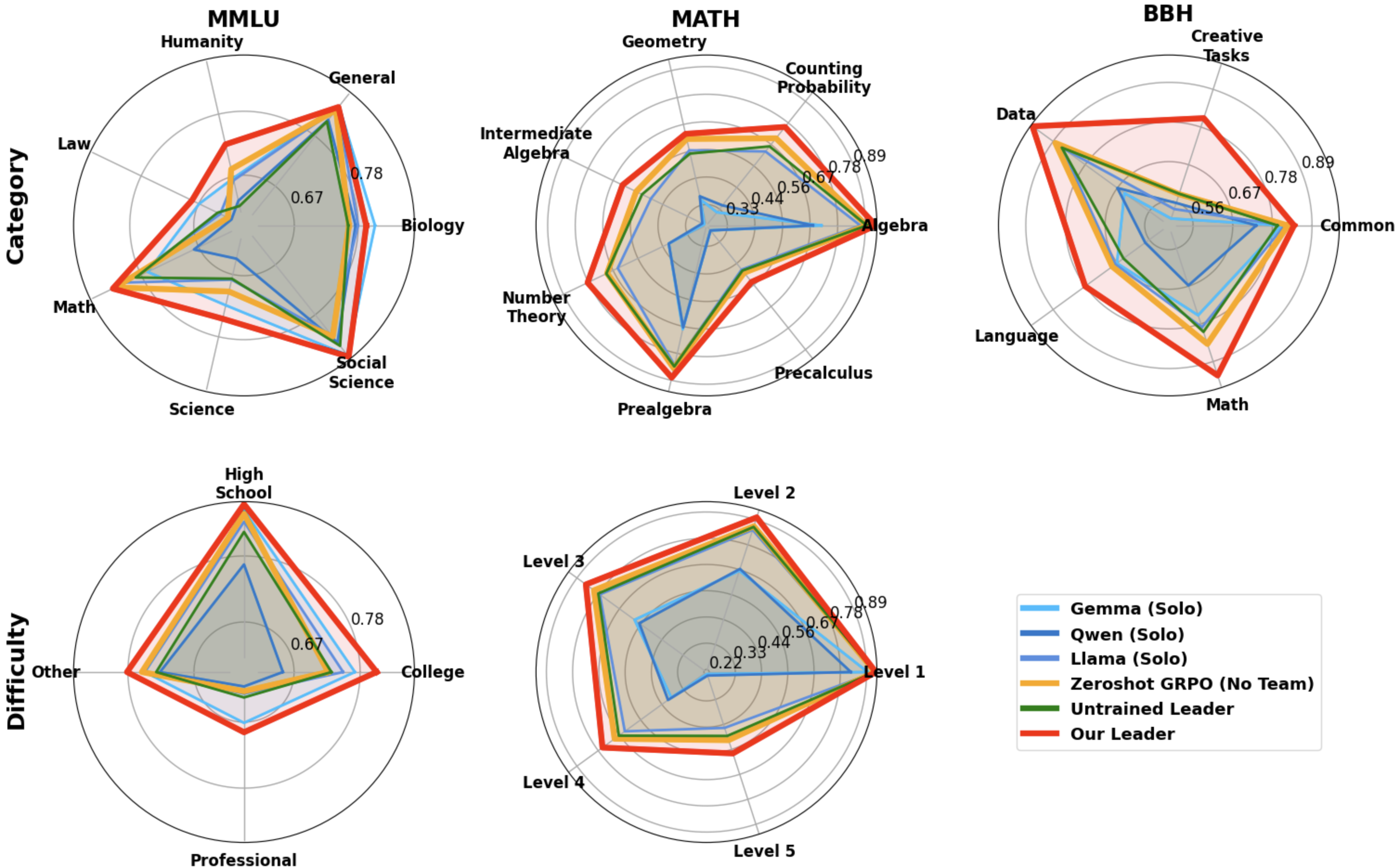}
    \caption{Our leader trained with MLPO, compared with an untrained leader, to zero-shot GRPO, and individual team performance, per category (top) and difficulty level (bottom) on MMLU (left) MATH (center) and BBH (right). Accuracy is reported after $5$ rounds of inference. Note that BBH does not have difficulty gradations.}
    \label{fig:MMLU_per_cata}
\end{figure}

\paragraph{\textbf{Leader vs Agent Team across Category and Difficulty}}
An ideal leader can leverage the unique strengths of the multi-agent team to produce solutions which are better than any one model could alone. 
Figure \ref{fig:MMLU_per_cata} shows the performance of our trained leader and each individual agent in the leader's team. 
Within these plots, we see two interesting observations.

First, the multi-agent pipeline with an untrained leader (green) struggles to surpass each of the individual agents (blue), both per-category and per-difficulty  across each of the benchmarks. 
In contrast, the multi-agent team with an \ours~trained leader can surpass the performance of each agent as well as the untrained leader in nearly all categories and difficulty (occasional ties with the best agent). 
Second, the model trained with zero-shot GRPO (orange) also struggles to dominate the untrained leader, as well as each agent, over categories and difficulties. 
The per-category and per-difficulty success of the zero-shot model is far more task-dependent than that of our trained leader. 
In particular, when examining MMLU, we see that the untrained agents or the untrained leader beats the zero-shot GRPO model on 4 our of 7 categories (Science, Social Science, Biology, and Law) and 2 out of 4 difficulties (professional and college). However, our trained leader dominates in all difficulties and 6 out of 7 categories (losing only on Biology), highlighting that the leader trained via our \ours~framework successfully acquires a balanced and robust skill set. \ours~consistently outperforms both the (trained) individual agents as well as the team with an untrained leader across almost all categories and difficulty levels. 

\paragraph{\textbf{Per-Round Improvement}}
\begin{figure}[]
\centering
\includegraphics[width=0.99\linewidth]{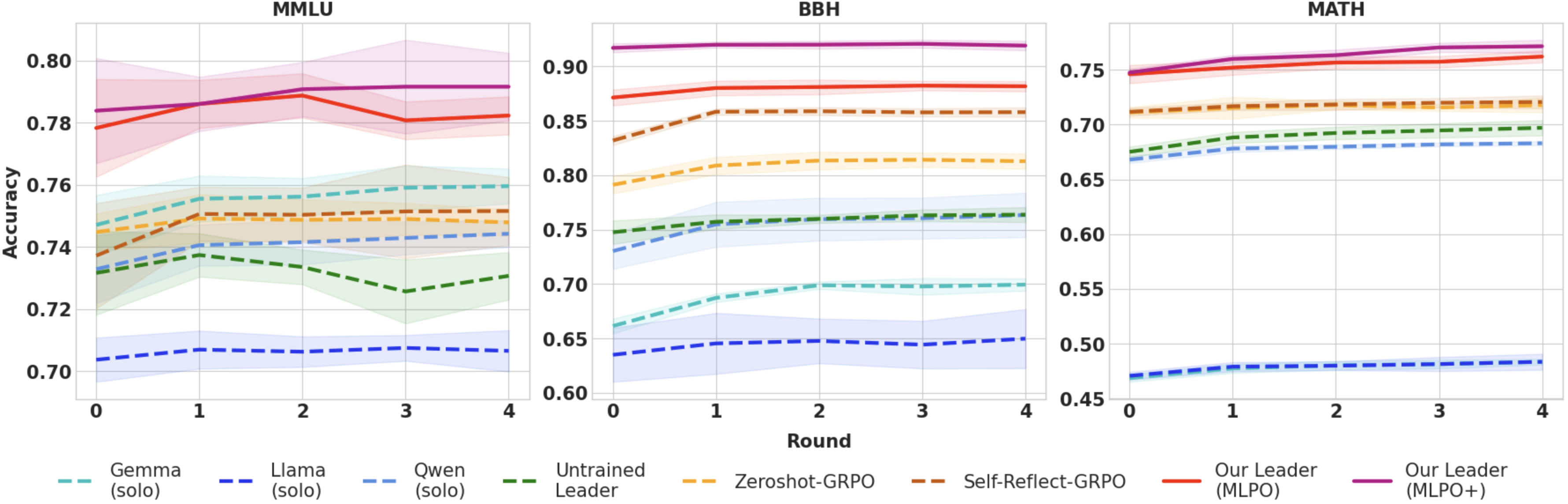}
\caption{Performance over multiple rounds of interaction. Shaded region represents $\pm2$ standard errors.}
\label{fig:perRound}
\end{figure}
Next, we examine the accuracy improvements achieved through iterative rounds of discussion between the leader and the agent team, as well as the self-reflection performance of individual agents (solo). Figure \ref{fig:perRound} (and previously shown Figure \ref{fig:majVote}) presents the accuracy progression for each method across these multiple rounds. 
Notably, even though our leader model is only trained on initial round data (round 0), we consistently observe slight accuracy improvements in subsequent rounds.
One possible explanation for this improvement is that in later rounds, agents benefit from additional information provided by the leader's earlier responses, enabling them to refine their own answers. This, in turn, may further assist the leader model in correcting its previous errors. However, further investigation is required to fully understand this phenomenon.

\paragraph{\textbf{Multi-Round Training with MLPO+}}
\label{subsec:mlpo_plus}
We now discuss a complementary phenomenon: the ability to further improve leader performance through additional training. 
We evaluate the performance of MLPO+, which fine-tunes the leader on agent-team responses from later rounds ($t > 0$), and find that it yields consistent improvements in leader performance.

The performance boost appears to stem from two complementary factors: first, the leader is trained on responses that better reflect those it will encounter during deployment, and second, it learns to better synthesize information from agent teams that have already been shaped by its prior guidance. 
Notably, these gains come with minimal additional computational cost, as the extra training data can still be collected offline.
Overall, these results highlight the value of iterative supervision: exposing the leader to more realistic interactions and more divers solutions during training increases the leader's efficacy.
\begin{table}[h]
    \centering
    \begin{tabular}{lccc}
            \toprule
            & MMLU & BBH & MATH\\
            \midrule
        MLPO &  0.782$_{\pm0.006}$ & 0.882$_{\pm0.005}$ &  0.762$_{\pm0.005}$ \\
         MLPO+ & 0.792$_{\pm 0.005}$ & 0.920$_{\pm 0.004}$ &  0.771$_{\pm 0.005}$\\
         \bottomrule
    \end{tabular}
    \caption{Results of a leader trained with MLPO and MLPO+ after 5 rounds of inference}
    \label{tab:MLPO_plus}
\end{table}

\begin{figure}[!htp]
    \centering
    \includegraphics[width=0.33\linewidth]{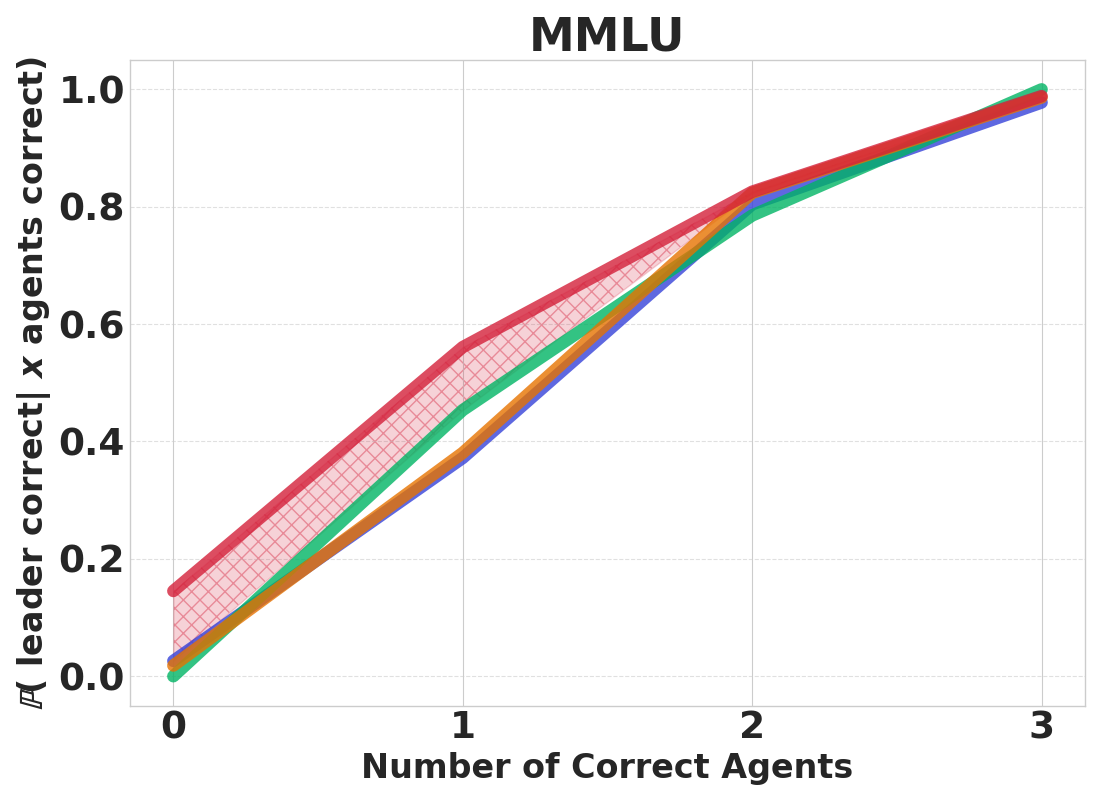}\includegraphics[width=0.33\linewidth]{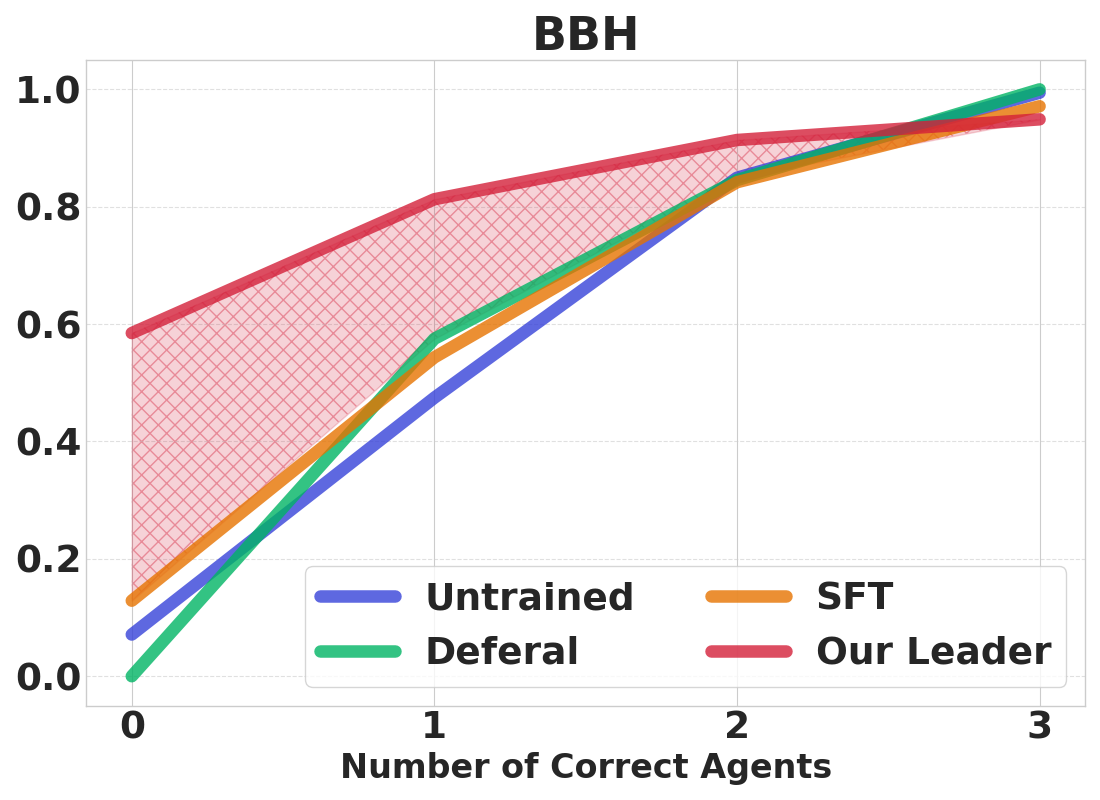}\includegraphics[width=0.33\linewidth]{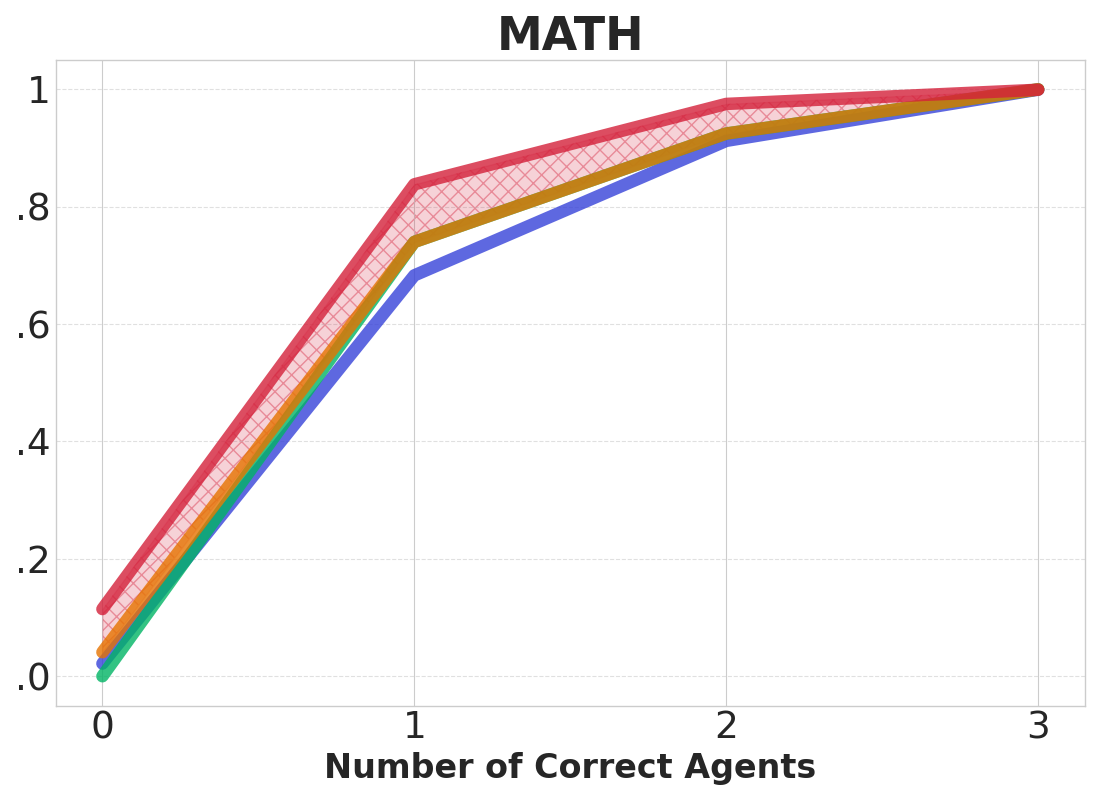}
    \caption{Leader performance conditioned on the number of correct agents in the team. 
    Shaded regions represent the degree to which our leader outperforms the next best method.
    }
    \label{fig:leaderCalPerAgent}
\end{figure}

\paragraph{\textbf{Leader Robustness to Agent Team's Ability}}
An effective leader should consistently produce correct solutions, by leveraging accurate responses from agents when available, and overriding incorrect ones when necessary. In Figure~\ref{fig:leaderCalPerAgent}, we compare the performance of various leaders as a function of agent correctness. 
To ensure a fair comparison, we report results only for round 0, where agent responses are not influenced by the leader. 
While all leaders perform well when all agents are correct, their accuracy drops substantially when agent correctness is low. 
Notably, our trained leader shows strong performance particularly when only few agents are correct, highlighting our method’s robustness to low team performance.
This effect is especially pronounced on the BBH dataset.

One possible explanation for this improved robustness could be better knowledge acquisition by the leader model during \ours, enabling it to independently generate correct answers without relying solely on the agent solutions at inference time. This hypothesis aligns with the observed increase in zero-shot accuracy as previously mentioned in Figure \ref{fig:zeroShotBar}. However, knowledge acquisition alone does not fully explain our leader’s performance. Specifically, we observe additional improvements when correct agent responses are available at inference time. This indicates that the leader has learned not only to independently solve problems but also to effectively evaluate the quality of agent inputs, discerning when to trust their advice and when to override incorrect team responses. 

\begin{figure}[h]
    \centering
    \includegraphics[width=0.99\linewidth]{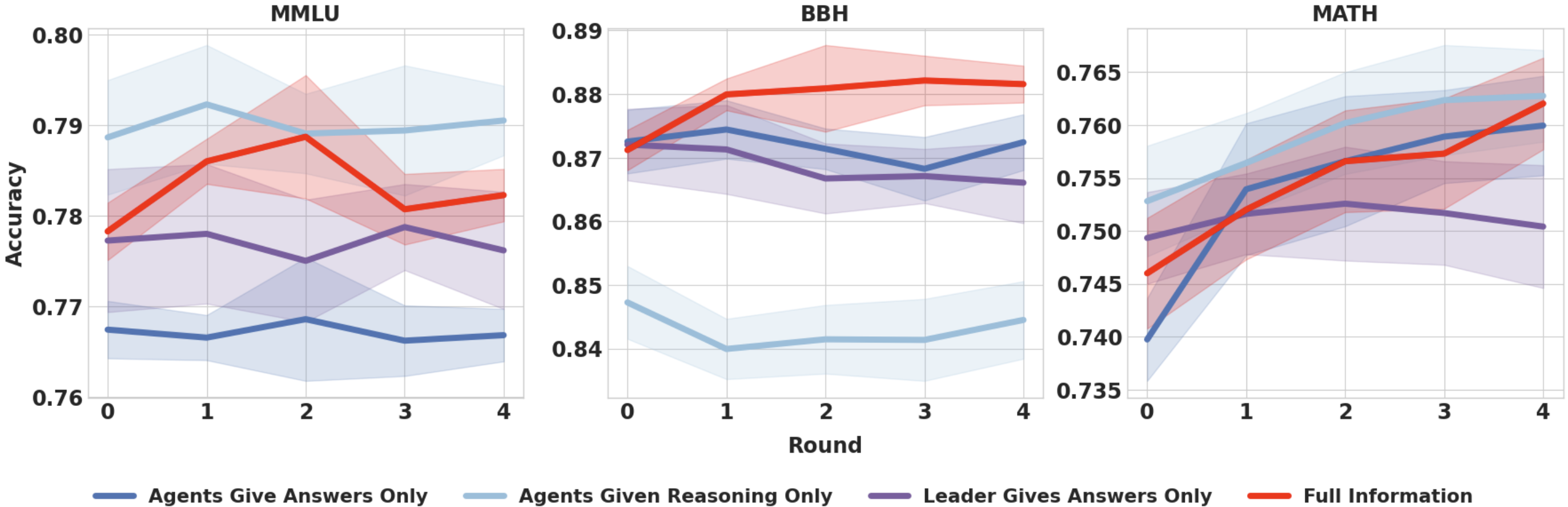}
    \caption{Leader performance over rounds of inference, varying by which information is shared between the agent team and the leader. Shaded region represents $\pm2$ standard errors.}
    \label{fig:whatTheLeaderSees}
\end{figure}

\paragraph{\textbf{Leader's Utilization of Agent Information}}
To better understand the mechanisms behind our trained leader’s superior performance, we analyze how it utilizes the information provided by the agent team. 
We conducted ablation experiments, selectively providing either agent reasoning, final answers, or both, to the leader during inference. The results of these experiments are summarized in Figure~\ref{fig:whatTheLeaderSees}.

Our findings consistently indicate that, in most scenarios, providing the leader with both the agents' reasoning processes and their final answers achieves the highest performance. 
This suggests that the leader effectively leverages the complete context provided by the agents, benefiting from both their detailed thought processes and their concluded solutions.

Interestingly, in the majority of cases, reasoning-only input achieves better performance than final-answer-only input. 
This implies that the leader primarily derives value from understanding the agents’ underlying reasoning rather than merely aggregating their final decisions. 
However, notable exceptions exist: for the BBH dataset, final-answer-only surpasses reasoning-only performance, although providing both components remains optimal. 
When deploying leaders trained with MLPO, utilizing agent full responses generally elicits the best leader performance. 

These observations indicate that the relative importance of reasoning versus final answers depends critically on both the dataset characteristics and the specific training regime. 
Including final answers may sometimes bias the leader towards incorrect conclusions, whereas reasoning-only input often encourages independent verification and robust inference. Nonetheless, overall, providing both reasoning and final answers to the leader consistently yields optimal or near-optimal results across diverse scenarios.

\begin{figure}[h]
    \centering
    \includegraphics[width=0.33\linewidth]{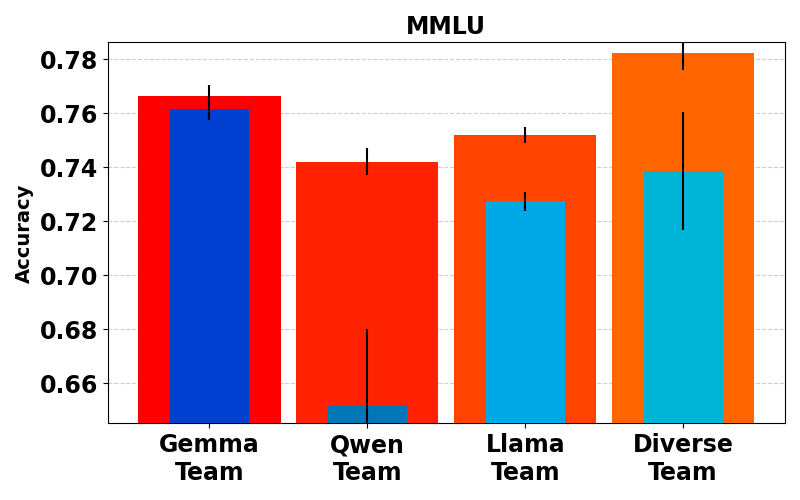}\includegraphics[width=0.33\linewidth]{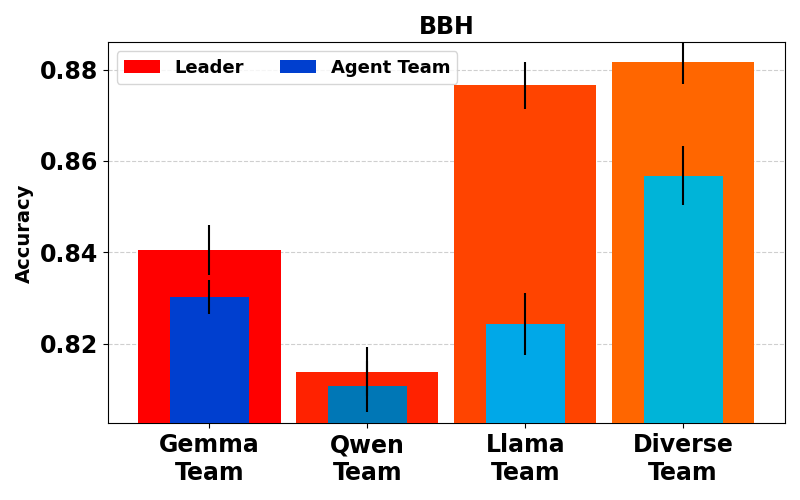}\includegraphics[width=0.33\linewidth]{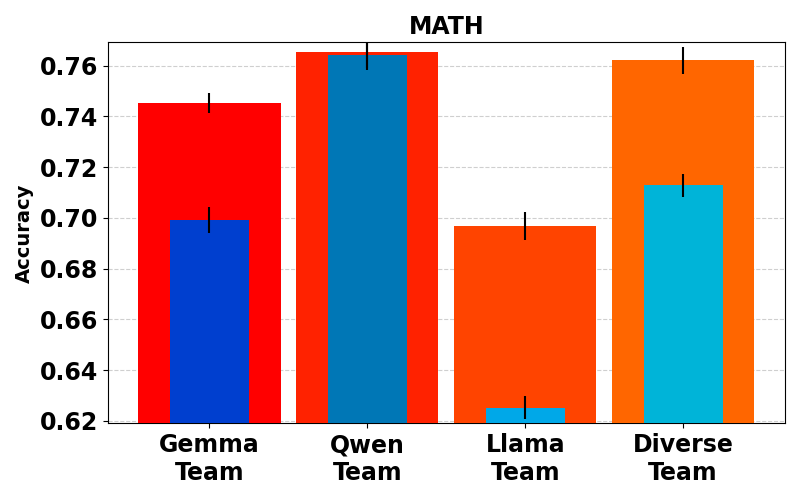}
    \caption{Leader accuracy (warm colors) and average team accuracy (cool colors) after 5 rounds of interaction.}
    \label{fig:teamComp}
\end{figure}

\paragraph{\textbf{Team Composition}}
To better understand the interaction dynamics between the leader and its team, we evaluate leader performance under different choices of agent teams; namely the diverse team used in our main results (Qwen-2.5, Llama-3.1, and Gemma-2), and homogeneous teams consisting of 3 copies of each model (e.g., 3 copies of Gemma-2). 
For each team configuration, a leader is trained specifically for that team. Figure~\ref{fig:teamComp} reports both the average leader accuracy and the average team accuracy after five rounds of interaction. 

Two key observations emerge from these results.
First, leaders paired with homogeneous teams exhibit more task-dependent performance. For example, on BBH dataset, the leader paired with Llama team achieves highest accuracy among all homogeneous teams, while on MATH, it achieves the lowest.
However, the leader paired with the heterogeneous team consistently matches or exceeds the performance of leaders paired with homogeneous teams. 
This highlights the benefit of deploying \ours~with a diverse agent team. 
Second, the gap between leader and team performance can be substantial, even after the team has had extensive interaction with the leader.
This underscores the leader’s critical role in providing high-quality solutions even when the underlying team struggles.

\subsection{Best Practices for Multi-Agent Guided Training}\label{sec:best_prac}
We conclude by highlighting key considerations in our multi-agent guided training paradigm that helped guide the design of our method.
Recall that during training, the leader observes the responses generated by the agent team before producing its own answer.
Thus, the leader is trained on prompt-response pairs of the form \texttt{(task + agent\_responses, leader\_response)}, in contrast to standard single-agent training, which relies solely on the task input.
This distinction gives rise to two natural questions:
\begin{enumerate}
    \item To what extent do the alternative solutions proposed by the agents contribute to the effectiveness of the trained leader?
    \item Given that the agent team can generate a multitude of different responses for a single task, how many sets of such response sets should the leader be exposed to during training for each task?
    \item Does filtering out easy questions improve overall training efficacy?
\end{enumerate}

\paragraph{\textbf{Importance of Alternative Solutions}}
To address the first question above, we assess the importance of alternative solutions provided by the agent team during both training and inference. 
As previously noted, the composition of the agent team, and thus the diversity (and quality) of solutions presented to the leader, can significantly affect performance. 
To investigate this, we evaluate five configurations that vary in terms of how alternative solutions are presented to the trained model (both at training and at inference time).

\begin{itemize}
\item \textit{GRPO Zeroshot Data}: No alternative solutions at training or inference; model sees only its own outputs.
\item \textit{SCoRe and GRPO on Self-Reflection Data}: Model is trained and inferenced with its own solutions.
\item \textit{Zeroshot as Leader}: No alternative solutions during training; observes agent team responses at inference.
\item \textit{Our Leader w. All Qwen-2.5 Team}: Trained and inferenced with homogeneous team of Qwen-2.5 agents.
\item \textit{Our Leader w. Diverse Team}: Trained and inferenced with heterogeneous team of three distinct agents.
\end{itemize}

Table~\ref{tab:round5_accuracy} presents the performance of each method after five rounds of inference.
Among these, the best performance is achieved by the leader trained, via our method, with responses from a diverse team of agents.
Interestingly, we observe no clear advantage when models are trained on their own solutions (as in SCoRe and GRPO Self-Reflect) compared to those trained on homogeneous agent team responses.
These results highlight the importance of sourcing alterative solutions form multiple distinct agents, which results in the highest-quality responses from the leader.

\begin{table}[]
\centering
\begin{sc}
\begin{tabular}{lccc}
\toprule
Method & MMLU & BBH & MATH \\\midrule
Untrained                & 0.734$_{\pm0.006}$ & 0.733$_{\pm0.010}$ & 0.666$_{\pm0.002}$ \\
\hline
GRPO Zeroshot Data       & 0.747$_{\pm0.003}$ & 0.814$_{\pm0.006}$ & 0.718$_{\pm0.006}$ \\
SCoRe                   & 0.752$_{\pm0.005}$ & 0.828$_{\pm0.004}$ & 0.721$_{\pm0.004}$ \\
GRPO Self-Reflect Data   & 0.762$_{\pm0.004}$ & 0.857$_{\pm0.006}$ & 0.720$_{\pm0.004}$ \\
Zeroshot as Leader & 0.742$_{\pm0.003}$ & 0.783$_{\pm0.002}$ & 0.729$_{\pm0.007}$ \\
\hline
Our Leader w. All Qwen-2.5 Team       & 0.742$_{\pm0.004}$ & 0.816$_{\pm0.008}$ & \textbf{0.767}$_{\pm0.003}$ \\
Our Leader w. Diverse Team  &                   \textbf{0.782}$_{\pm0.006}$ & \textbf{0.882}$_{\pm0.005}$ & 0.762$_{\pm0.005}$ \\
\bottomrule
\end{tabular}
\end{sc}
\caption{Accuracy after 5 rounds of inference. 
}
\label{tab:round5_accuracy}
\end{table}

\paragraph{\textbf{Leader's Exposure to Distinct Agent Solutions Per Task During Training}} We next examine how the number of distinct solution sets per task provided to the leader during training affects performance. We use BBH as an exemplar for this ablation.
Specifically, we train the leader using 1, 4, and 8 sets of agent responses per task (in our main experiments, we adopt 4 sets per task). 
To ensure a fair comparison, all variants use the same total number of training examples. Table~\ref{tab:solution_sets} reports leader accuracy under each setting.
We observe that exposing the leader to multiple solution sets improves performance, though gains begin to plateau beyond 4 sets. 
This suggests that diverse agent responses help the leader generalize to varied team behaviors, but that additional diversity beyond a certain point offers diminishing returns.

\begin{table}[h]
\centering
\begin{sc}
\begin{tabular}{ccc}
\toprule
\multicolumn{3}{c}{Number of Solution Sets Seen at Training} \\
\midrule
1 Set & 4 Sets & 8 Sets \\
\midrule
0.890$_{\pm 0.005}$ & 0.917$_{\pm 0.003}$ & 0.906$_{\pm 0.002}$ \\
\bottomrule
\end{tabular}
\end{sc}
\caption{Leader accuracy for different numbers of solution sets per task during training (BBH), $\pm2$ standard errors.}
\label{tab:solution_sets}
\end{table}

We also examine the impact of training-time task difficulty on leader performance.
Specifically, we consider how ``easy'' a given task is for the agent team, independent of the leader. Any task where at least 75\% of the agent responses are correct is removed from the training data. Training on these filtered tasks leads to improved leader performance at test time.
One possible explanation is that including tasks on which the agent team already performs well may cause the leader to become overly reliant on agent responses, reducing its incentive to critically evaluate or override incorrect solutions.

\begin{table}[h]
\centering
\begin{sc}
\begin{tabular}{cc}
\toprule
unfiltered training & filtered training \\
\midrule
0.869$_{\pm 0.004}$ & 0.882$_{\pm0.005}$ \\
\bottomrule
\end{tabular}
\end{sc}
\caption{Accuracy of leader trained with and without difficulty-based filtering for BBH, with $\pm2$ standard errors.}
\label{tab:solution_filtering}
\end{table}

It is worth noting these ablations cannot be directly compared to the results in the main body as we use a different amount of training data as to ensue fair comparisons between the different ablations.

\begin{figure}[h]
    \centering
    \includegraphics[width=0.75\linewidth]{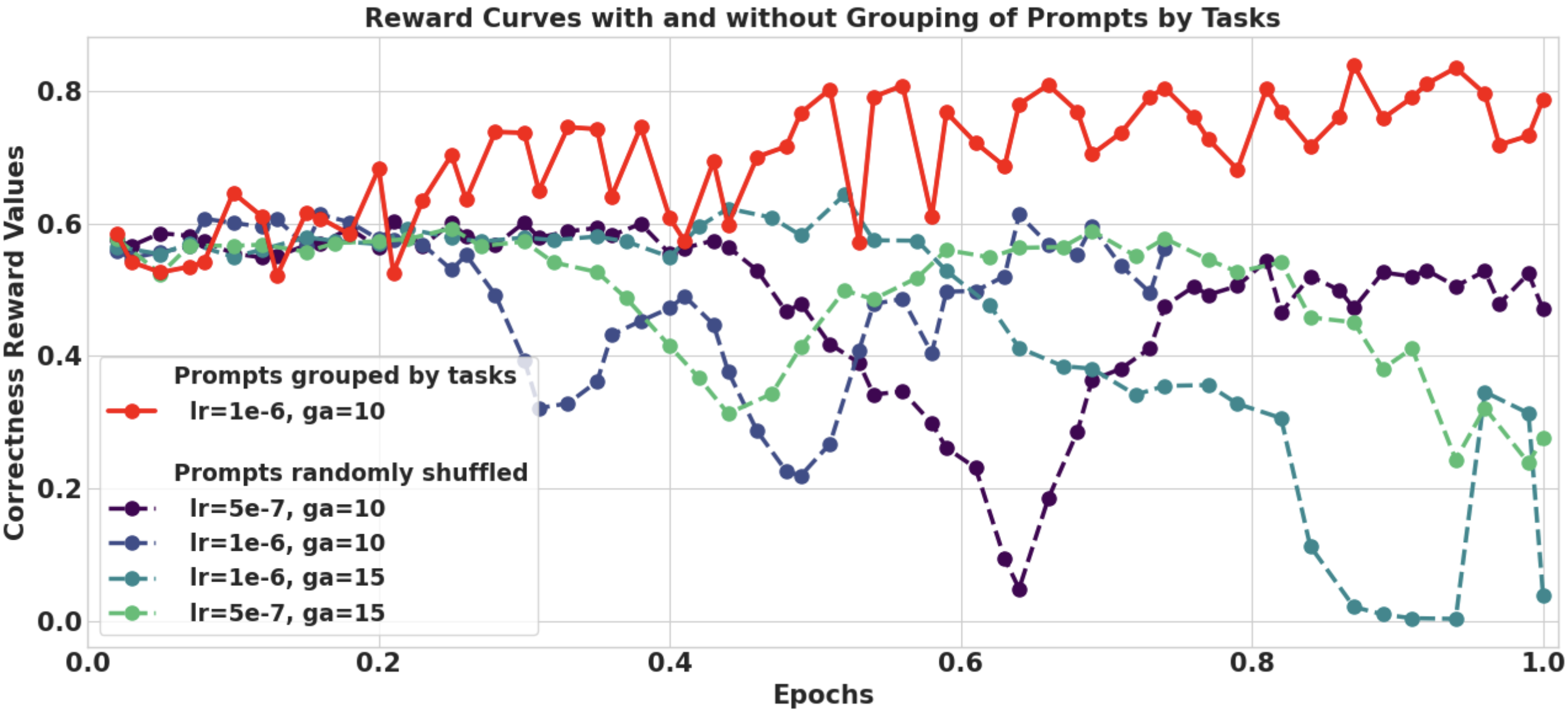}
    \caption{Reward curves for training with grouped solution sets vs randomly shuffled solution sets. Learning rate and gradient accumulation steps are denoted respectively as \textit{lr} and \textit{ga}.}
    \label{fig:reward}
\end{figure}

\paragraph{\textbf{Solution Set Ordering}}
Next, we discuss the ordering in which the leader is exposed to the different sets of solutions. 
For our main experiments, we sample four sets of agent solutions for each task $x$.
In Figure \ref{fig:reward} we see difference in reward curves for training with shuffled solution sets, and grouped solution sets (i.e., the leader sees all four solution sets in a single batch).
We find that grouping solutions by task makes training much more stable and results in higher reward for the trained leader.

\paragraph{\textbf{Alterative Choice of Leader}}
Lastly we remark on our choice of model for the trained leader. 
Throughout the main body, we present results when using Qwen-2.5 as the leader. 
In Appendix \ref{sec:sup_diff_leader} we present our main set of results for alterative leaders. 
We observe similar results, utilizing our multi-agent guided training results in the highest efficacy, although we observe less performance increase using Gemma-2 and Llama-3.1 as the leader.
This suggested that the most effective leaders may be models capable of more advanced reasoning, such as Qwen-2.5.
This matches with intuition as an effective leader must be skilled at reasoning through the agent responses.

\section{Conclusion}
We introduce a hierarchical multi-agent framework leveraging our proposed Multi-Agent Guided Leader Optimization (\ours) method, specifically designed to enhance the collaborative reasoning capabilities of large language models (LLMs). By explicitly training only a single leader model, our approach introduces novel multi-agent guided training and inference techniques, effectively integrating diverse insights from a team of untrained peer agents. Extensive empirical results on challenging benchmarks demonstrate significant performance improvements over single-agent and other multi-agent baselines. 

Nevertheless, our method also has certain limitations. These include increased context lengths required both during training and inference, higher computational demands at inference time for optimal performance, and reduced parallelizability due to sequential leader-agent interactions. 
Addressing these limitations could involve exploring strategies like selective agent querying and caching to reduce training/inference load as well as different agent team combinations to reduce the number of rounds needed.
Despite these constraints, our framework remains highly adaptable. It supports both parallel and sequential scaling strategies as needed, and notably, achieves strong performance even when deployed without the agent team at inference time.

\bibliographystyle{plainnat}  
\bibliography{main}

\newpage
\appendix
\section*{Appendix}

\section{Different Choice of Leader }\label{sec:sup_diff_leader}
In the main body we use Qwen-2.5-7B-Instruct as our leader. 
In Tables \ref{tab_app:benchmark-results} and \ref{tab:AllBenchmarksLlama}, we present results when Gemma-2-9B-it and Llama-3.1-8B-Instruct are used as the leader agent, respectively. 

To ensure a fair comparison against baselines when varying the leader agent in our method, we alternate the ``main" model used in the baselines (e.g., for SelectLLM the trained model which makes the selection is the same model our leader). 
For any method that uses a team (e.g., MAD) we keep the team the same.

\begin{table}[htb]
\centering
\renewcommand{\arraystretch}{1.2}
\setlength{\tabcolsep}{12pt}
\begin{sc}
\begin{tabular}{llccc}
\toprule
Type & Method & MMLU & BBH & MATH \\
\midrule
\multirow{5}{*}{\shortstack[l]{Training-\\free}} 
& Zero-Shot & 0.748$_{\pm0.006}$ & 0.659$_{\pm0.005}$ & 0.468$_{\pm0.002}$ \\
& Self-Reflect \cite{madaan2023self} & 0.763$_{\pm0.005}$ & 0.697$_{\pm0.005}$ & 0.481$_{\pm0.003}$ \\
& MAD (Div) \cite{du2023improving} & 0.771$_{\pm0.002}$ & 0.785$_{\pm0.002}$ & 0.653$_{\pm0.009}$ \\
& MAD (Hom) \cite{du2023improving} & 0.776$_{\pm0.007}$ & 0.717$_{\pm0.012}$ & 0.496$_{\pm0.008}$ \\
& Untrained Leader & 0.738$_{\pm0.004}$  & 0.759$_{\pm0.005}$  &  0.655$_{\pm 0.010}$ \\
\hline
\multirow{6}{*}{\shortstack[l]{Training-\\based}} 
&
ACC-Collab \cite{acccollab} & 0.786$_{\pm0.018}$ & 0.742$_{\pm0.007}$ & 0.501$_{\pm0.006}$ \\
& SelectLLM \cite{maurya2024selectllm} & 0.764$_{\pm0.001}$ & 0.751$_{\pm0.001}$ & 0.549$_{\pm0.001}$ \\
& SCoRe \cite{kumar2024training}  & 0.755$_{\pm0.005}$ & 0.688$_{\pm0.007}$ & 0.452$_{\pm0.003}$ \\
& Deferal Leader & 0.741$_{\pm0.005}$ & 0.759$_{\pm0.007}$ &0.701$_{\pm0.003}$ \\
& Zero-Shot GRPO \cite{grpo} & 0.747$_{\pm0.004}$ & 0.680$_{\pm0.003}$ & 0.440$_{\pm0.003}$ \\
& Self-Reflect GRPO \cite{grpo} + \cite{madaan2023self} & 0.772$_{\pm0.002}$ & 0.764$_{\pm0.005}$ & 0.500$_{\pm0.004}$ \\
\hline
Ours & MLPO Leader & 0.750$_{\pm0.003}$ & 0.813$_{\pm0.003}$ & 0.697$_{\pm0.012}$ \\
\bottomrule
\end{tabular}
\end{sc}
\caption{Accuracy of each method on MMLU, BBH, and MATH benchmarks when Gemma-2-9B-it is used as the leader. All baselines are implemented with Gemma-2-9B-it. The leader model trained with MLPO is denoted as Our Leader.}
\label{tab_app:benchmark-results}
\end{table}

\begin{table}[htb]
\centering
\renewcommand{\arraystretch}{1.2}
\setlength{\tabcolsep}{12pt}
\begin{sc}
\begin{tabular}{llccc}
\toprule
Type & Method & MMLU & BBH & MATH \\
\midrule
\multirow{5}{*}{\shortstack[l]{Training-\\free}} 
& Zero-Shot & 0.703$_{\pm0.005}$ & 0.638$_{\pm0.014}$ & 0.470$_{\pm0.003}$ \\
& Self-Reflect \cite{madaan2023self} & 0.712$_{\pm0.006}$ & 0.656$_{\pm0.014}$ & 0.484$_{\pm0.005}$ \\
& MAD (Div) \cite{du2023improving} & 0.771$_{\pm0.002}$ & 0.785$_{\pm0.002}$ & 0.653$_{\pm0.009}$ \\
& MAD (Hom) \cite{du2023improving} & 0.734$_{\pm0.007}$ & 0.707$_{\pm0.015}$ & 0.533$_{\pm0.009}$ \\
& Untrained Leader & 0.714$_{\pm0.006}$  &  0.741$_{\pm0.005}$ & 0.662$_{\pm 0.014}$ \\
\hline
\multirow{6}{*}{\shortstack[l]{Training-\\based}} 
&
ACC-Collab \cite{acccollab} & 0.755$_{\pm0.020}$ & 0.740$_{\pm0.018}$ & 0.485$_{\pm0.004}$ \\
& SelectLLM \cite{maurya2024selectllm} & 0.766$_{\pm0.001}$ & 0.762$_{\pm0.001}$ & 0.565$_{\pm0.001}$ \\
& SCoRe \cite{kumar2024training} & 0.524$_{\pm0.006}$ & 0.638$_{\pm0.009}$ & 0.325$_{\pm0.006}$ \\
& Deferal Leader & 0.732$_{\pm0.005}$ & 0.752$_{\pm0.007}$ & 0.692$_{\pm0.004}$ \\
& Zero-Shot GRPO \cite{grpo} & 0.714$_{\pm0.006}$ & 0.712$_{\pm0.007}$ & 0.466$_{\pm0.006}$ \\
& Self-Reflect GRPO \cite{grpo} + \cite{madaan2023self} & 0.723$_{\pm0.006}$ & 0.721$_{\pm0.007}$ & 0.487$_{\pm0.006}$ \\
\hline
Ours & MLPO Leader & 0.741$_{\pm0.008}$ & 0.771$_{\pm0.011}$ & 0.720$_{\pm0.007}$ \\
\bottomrule
\end{tabular}
\end{sc}
\caption{Accuracy of each method on MMLU, BBH, and MATH benchmarks when Llama-3.1-8B-Instruct is used as the leader. All baselines are implemented with Llama-3.1-8B-Instruct. The leader model trained with MLPO is denoted as Our Leader.}
\label{tab:AllBenchmarksLlama}
\end{table}

\section{Baseline Details}
Here we outline our setup for training baselines.

\paragraph{\textbf{MAD}} Deployed with Qwen 2.5 7B Instruct, Llama 3.1 8B Instruct, and Gemma 2 9B Instruct \textbf{MAD (Div)}, and three copies of the same model (e.g., 3 copies of Qwen 2.5 7B Instruct) \textbf{MAD (Hom)}.

\paragraph{\textbf{Zeroshot-GRPO}} Trained with Qwen 2.5 7B Instruct for main results, and either Llama 3.1 8B Instruct, or Gemma 2 9B Instruct for Section \ref{sec:sup_diff_leader} (selecting the same model as the leader trained with MLPO). We use a learning rate of between 1e-6 to 5e-7, KL-regularization of $\beta = 0$.

\paragraph{\textbf{ACC-Collab}} Trained two copies of Qwen 2.5 7B Instruct (one trained as the actor, one as the critic) for the main results. For results in Section \ref{sec:sup_diff_leader} we use two copies of either Llama 3.1 8B Instruct or Gemma 2 9B Instruct (selecting the same model as the leader trained with MLPO). 
We train with RPO using learning rates between 1e-6 to 5e-6, a regularization term of $\alpha_{\text{RPO}} \in \{1.3, 0.7, 1\}$.

\paragraph{\textbf{SelectLLM}} The selection LLM is trained by taking Qwen 2.5 7B Instruct for main results (and Llama 3.1 8B Instruct, or Gemma 2 9B Instruct for Section \ref{sec:sup_diff_leader}) and retrianing the last two fully connected layers. We use a learning rate of 1e-5 with weight decay of 0.01. 
We deploy the variant which makes model selections via maximum confidence score, with a majority voting budget of 20 (keeping with the 20 total inference-time generations used by our pipeline for 5 rounds of inference). The selection model has access to all three agents: Qwen 2.5 7B Instruct, Llama 3.1 8B Instruct, and Gemma 2 9B Instruct.

\paragraph{\textbf{SCoRe}} Trained with Qwen 2.5 7B Instruct for main results, and either Llama 3.1 8B Instruct, or Gemma 2 9B Instruct for Section \ref{sec:sup_diff_leader} (selecting the same model as the leader trained with MLPO). We use a learning rate of between 1e-6 to 5e-6, a \emph{bonus scale} of $\alpha=10$, stage one regularization of $\beta_1 = 0.01$, and a stage two regularization of $\beta_2 = 0.1$. In stage two we use four samples of updated answers per prompt and group these answers into the same batch (keeping with the same data grouping utilized by MLPO as outlined in Section \ref{sec:multi_agent_training}).

\paragraph{\textbf{Deferral Leader}}  The leader is trained by taking Qwen 2.5 7B Instruct for main results (and Llama 3.1 8B Instruct, or Gemma 2 9B Instruct for Section \ref{sec:sup_diff_leader}) and retrianing the last two fully connected layers. The deferral leader views all solutions from the agent team and select one solution.
We use a learning rate of 1e-5 with weight decay of 0.01. 
The deferral leader is trained on the same data that the MLPO leader is trained on (outlined in Section  \ref{sec:multi_agent_training}), and is deployed on the same team as the MLPO leader.

\section{Prompts Used in Our Pipeline}
In this section, we present the prompts used in our pipeline, with examples drawn specifically from the MATH dataset. Analogous prompt structures are used for the other datasets.

\subsection{Agent Prompts}
We use structured and context-aware prompts for the agent models, both during inference and while collecting the training data. The prompts differ slightly depending on the inference round.

\paragraph{\textbf{Round 0 Prompt}}
In the first round, agents receive the question along with a standard solving plan. The prompt encourages step-by-step reasoning, adherence to factual premises, and concludes with a boxed answer.

\begin{tcolorbox}[title=Agent Prompt for Round 0, breakable]
\ttfamily
You are part of a team of 3 LLMs responsible for solving a math problem. 
You are Agent\_[agent\_number]. Strictly follow the given plan to solve the following math problem by thinking step by step through the plan. 
Make sure to first define the premises of the question, and make sure that the answer is consistent with the premises. 
Strictly stick to the facts and question provided. \\

*** Question: [question text] *** \\

*** Plan: Solve the given math problem. Think step-by-step, providing detailed calculations and reasoning for your steps. *** \\

Regardless of the approach, always conclude with: \\

\texttt{Therefore, the final answer is: \$\textbackslash boxed\{[answer]\}\$}. 
Where [answer] is just the final number or expression that solves the problem.
\end{tcolorbox}

\paragraph{\textbf{Subsequent Round Prompts}}
In later rounds, agents are additionally provided with their previous solution and feedback from the aggregator (leader). This guides the agent to reflect, revise, and correct any issues based on the leader’s evaluation and questions.

\begin{tcolorbox}[title=Agent Prompt for Subsequent Rounds, breakable]
\ttfamily
You are part of a team of 3 LLMs collaborating to solve a math problem. 
You are Agent\_[agent\_number]. Your goal is to improve your previous response using feedback in the form of questions.
Strictly follow the plan and think step-by-step through the math problem. 
Make sure your answer is consistent with the premises and addresses the raised questions thoroughly. \\

*** Question: [question text] *** \\

Your previous solution:
[previous\_solution] \\

Additionally, the aggregator has evaluated agent responses from the previous round and has also raised questions about your previous response. \\

The aggregator's output:
[aggregator\_response] \\

*** Plan: \\

1. Carefully reflect on the aggregator's feedback and your previous solution.  \\

2. Revise your answer step-by-step to improve its correctness and clarity. Address each question raised where relevant.  \\

3. Double-check for any logical, calculation, or reasoning errors. \\ \\

Regardless of the approach, always conclude with: \\

\texttt{Therefore, the final answer is: \$\textbackslash boxed\{[answer]\}\$}. 
Where [answer] is just the final number or expression that solves the problem.
\end{tcolorbox}

\subsection{Leader Prompts}
The leader model is prompted to evaluate the responses of all agents to the given problem and produce a final answer. The prompt instructs the leader to critically analyze the agents’ outputs, identify correct reasoning or mistakes, and then synthesize a coherent and accurate solution. The leader is explicitly instructed to produce two structured blocks: a reasoning trace enclosed in \texttt{<think>} tags, and a final answer in the \texttt{<answer>} block, which must conclude with a boxed expression.

\begin{tcolorbox}[title=Leader Prompt, breakable]
\ttfamily
You are an expert aggregator LLM tasked with evaluating multiple agents' responses to a math problem. Your goal is to critically analyze all agent responses, identify correct reasoning or errors, and then provide a unified answer. \\ 

Question: [question text] \\ 

Agent 1 Response: [agent\_1 response] \\

Agent 2 Response: [agent\_2 response] \\

Agent 3 Response: [agent\_3 response] \\ \\

Please complete the following two blocks in order: \\

1. \texttt{<think>...</think>}: A long, detailed chain-of-thought reasoning process. \\

2. \texttt{<answer>...</answer>}: Your final answer should be aggregated from the best elements of the agents’ responses. \\ \\

- End the answer with: \texttt{Therefore, the final answer is: \$\textbackslash boxed\{[answer]\}\$}.
\end{tcolorbox}

\subsection{Generating Backtracking Data for SFT}

To teach the leader model natural backtracking and self-correction behavior, we construct a supervised fine-tuning (SFT) dataset comprising synthetic completions that mimic realistic reasoning failures followed by recovery. This is inspired by recent work encouraging naturalistic self-correction patterns in LLMs \cite{qin2024o1,guo2025deepseek,team2025kimi}.

We begin by running our untrained multi-agent pipeline to collect 16 distinct completions from the leader for each agent team generation. We discard examples where all completions are incorrect. For the remaining examples, we randomly select one correct and one incorrect leader completion. These are then used to construct a prompt that asks an untrained leader model to imitate a reasoning process that starts with plausible but flawed logic, then backtracks and self-corrects to reach the correct answer. The generated backtracked completions form the core of our SFT dataset.

Below, we include the exact prompt used to generate such backtracking examples for the MATH dataset.

\begin{tcolorbox}[title=Backtracking Generation Prompt (for MATH), breakable]

\ttfamily

You are an AI aggregator tasked with evaluating multiple agent responses to a question and aggregating them into a coherent and accurate final answer. You will be provided with the original prompt of the aggregator, a correct aggregation (which evaluated the agents correctly and arrived at the correct final answer), and an incorrect aggregation (which arrived at an incorrect final answer). Your goal is to mimic an aggregator which tries to aggregate the agent responses, first makes mistakes in its reasoning process by going down the wrong path and then backtracks to the correct reasoning. \\

The task consists of the following components: 

1. **Prompt**: The original math question and agent responses.

2. **Incorrect aggregator response**: Response from a previous aggregator that contains errors in aggregation or reasoning and hence arrives at an incorrect final answer. These should be carefully reviewed for mistakes.  

3. **Correct aggregator response**: Response from a previous aggregator that contains the correct reasoning, evaluation and aggregation.  \\
 
Here’s the input data for your aggregation:

Question:  
[question text]

Agent 1 response:  
[agent\_1 response]

Agent 2 response:  
[agent\_2 response]

Agent 3 response:  
[agent\_3 response]

Incorrect reasoning from Previous Aggregator:  
[extracted <think> block from incorrect leader response]

Correct reasoning from Previous Aggregator:  
[extracted <think> block from correct leader response] \\

Here’s your task breakdown:

- You should start by examining the agent responses and considering the provided aggregator responses.  

- **IMPORTANT**: You should use the incorrect aggregator response to mimic an aggregator which makes mistakes in its reasoning process before backtracking in your reasoning, acknowledging the mistakes, and correcting your approach based on the correct aggregator response provided.  

\quad  - As you work through the problem, narrate your thinking naturally just like a student working things out on paper. When you notice a mistake, acknowledge it with a natural reflection like: 

\qquad    - "Wait, that doesn't seem right."  

\qquad    - "Hmm, I think I made a mistake there." 

\qquad    - "Actually, let me go back and check that."  

\qquad    - "Maybe I should rethink this."

Your solution should feel like a continuous, thoughtful monologue that includes:  \\
- One or more initial (realistic) mistakes in reasoning.  \\
- Natural moments of self-correction and backtracking.  \\
- Accurate evaluation of agent correctness (as provided in the correct response).  \\
- A correct final solution (as provided in the correct response).  \\

At the end of your response you should include the final answer as:  
\texttt{Final answer: \$\textbackslash boxed\{[answer]\}\$}

**IMPORTANT**: Your final answer must be consistent with the correct aggregator's final answer.

\end{tcolorbox}

\paragraph{\textbf{Example Backtracking Generation}}

Below, we include an example used in our SFT dataset generation. The first box shows the prompt, which includes the agent team's responses for a given MATH question, and the second box contains the backtracked response generated using the methodology outlined above, which is used as the completion to SFT the leader for the given prompt below.

\vspace{0.5em}

\begin{tcolorbox}[title=Example Prompt with Task + Agent Team Generation, breakable]

\ttfamily
You are an expert aggregator LLM tasked with evaluating multiple agents' responses to a math problem. Your goal is to critically analyze all agent responses, identify correct reasoning or errors, and then provide a unified answer. \\

Question: Alex needs to catch a train. The train arrives randomly some time between 1:00 and 2:00, waits for 10 minutes, and then leaves. If Alex also arrives randomly between 1:00 and 2:00, what is the probability that the train will be there when Alex arrives? \\

\textbf{Agent 1 Response:} **Problem Definition**

To solve this problem, we need to understand the conditions for Alex to catch the train. The train arrives randomly between 1:00 and 2:00, waits for 10 minutes, and then leaves. Alex also arrives randomly between 1:00 and 2:00. \\

**Step 1: Define the Time Interval**

Let's define the time interval from 1:00 to 2:00 as a 60-minute window. We can represent this interval as a continuous segment from 0 to 60 minutes, where 0 represents 1:00 and 60 represents 2:00. \\

**Step 2: Calculate the Total Time Interval**

The total time interval for both the train's arrival and Alex's arrival is 60 minutes. \\

**Step 3: Calculate the Train's Waiting Time**

The train waits for 10 minutes after its arrival. This means that if the train arrives at time `t', it will be present at the station from `t' to `t + 10' minutes. \\

**Step 4: Define the Conditions for Alex to Catch the Train**

For Alex to catch the train, he must arrive during the 10-minute waiting period after the train's arrival. This means that Alex's arrival time 'A' must satisfy the condition:

$1 \leq A \leq t + 10$, where $t$ is the train's arrival time. \\

**Step 5: Calculate the Probability**

To calculate the probability, we need to find the area of the region where Alex arrives while the train is present. This can be visualized as a shaded area in a 2D coordinate system, where the x-axis represents the train's arrival time and the y-axis represents Alex's arrival time.

The total area of the sample space is a square with side length 60, representing the total time interval.

The region where Alex catches the train is a trapezoid with bases 60 and 50 and height 10. \\

**Step 6: Calculate the Area of the Trapezoid**

The area of a trapezoid is given by:

Area = (1/2) × (base1 + base2) × height
= (1/2) × (60 + 50) × 10
= (1/2) × 110 × 10
= 550 \\

**Step 7: Calculate the Total Area of the Sample Space**

The total area of the sample space is a square with side length 60, so the area is:

Total Area = 60 × 60
= 3600 \\

**Step 8: Calculate the Probability**

The probability of Alex catching the train is the ratio of the area of the trapezoid to the total area of the sample space:

Probability = Area of Trapezoid / Total Area
= 550 / 3600
= 55/360
= 11/72 \\

**Conclusion**

Therefore, the final answer is: $\boxed{\frac{11}{72}}$. \\ \\

\textbf{Agent 2 Response:} Here's a step-by-step solution to the problem:

**1. Define the Sample Space:**

The sample space consists of all possible pairs of arrival times for the train and Alex.  Since both can arrive anytime between 1:00 and 2:00, we have a range of 60 minutes for each. \\

**2.  Visualize the Problem:**

Imagine a square where the x-axis represents the train's arrival time and the y-axis represents Alex's arrival time. Each point within the square represents a possible combination of arrival times. \\

**3.  Identify Favorable Outcomes:**

We want the cases where Alex arrives within the 10-minute window when the train is waiting.  This forms a 10-minute wide strip along the line where the train's arrival time is equal to Alex's arrival time minus 10 minutes. \\

**4. Calculate Areas:**

* **Total Area:** The total area of the square represents all possible outcomes, which is 60 minutes * 60 minutes = 3600 square minutes.

* **Favorable Area:** The favorable area is a 60-minute long strip, with a width of 10 minutes. So, the favorable area is 60 minutes * 10 minutes = 600 square minutes. \\

**5. Calculate Probability:**

Probability is the ratio of favorable outcomes to total outcomes:

Probability = (Favorable Area) / (Total Area) = 600 square minutes / 3600 square minutes = 1/6

Therefore, the final answer is: $\boxed{\frac{1}{6}}$. \\ \\

\textbf{Agent 3 Response:} To solve this problem, we need to determine the probability that Alex will arrive at the train station during the time the train is present. We can visualize this problem using a coordinate plane where both the train's arrival time and Alex's arrival time are plotted. \\

\#\#\# Step 1: Define the time intervals

- The train arrives at a random time between 1:00 and 2:00. We can represent this time interval as \( [0, 60] \) minutes past 1:00.
- Similarly, Alex arrives at a random time between 1:00 and 2:00, which we can also represent as \( [0, 60] \) minutes past 1:00. \\

\#\#\# Step 2: Set up the coordinate system

We can plot the arrival times of the train and Alex on a coordinate plane where the x-axis represents the train's arrival time and the y-axis represents Alex's arrival time. Both axes range from 0 to 60 minutes. \\

\#\#\# Step 3: Define the region of interest

The train waits for 10 minutes after its arrival, so the train will be at the station from \( t \) to \( t + 10 \), where \( t \) is the train's arrival time. Alex will catch the train if he arrives during this 10-minute window.  \\

\#\#\# Step 4: Identify the favorable region

We need to find the area on the coordinate plane where Alex arrives during the train's 10-minute window. This area can be visualized as a band that runs horizontally from \( t \) to $t+10$ for each \( t \) from 0 to 50 (since if the train arrives at 50 minutes past 1:00, it will leave at 60 minutes past 1:00, and Alex can still catch it if he arrives at 60 minutes). \\

\#\#\# Step 5: Calculate the area of the favorable region

The favorable region is a band that is 10 units wide and 50 units long (from 0 to 50 on the x-axis). The area of this region is:
\[ \text{Area of favorable region} = 10 \times 50 = 500 \text{ square units} \] \\

\#\#\# Step 6: Calculate the total possible region

The total possible region is a square with side length 60, so the area is:
\[ \text{Area of total region} = 60 \times 60 = 3600 \text{ square units} \] \\

\#\#\# Step 7: Determine the probability

The probability that Alex will catch the train is the ratio of the area of the favorable region to the area of the total region:
\[ \text{Probability} = \frac{\text{Area of favorable region}}{\text{Area of total region}} = \frac{500}{3600} = \frac{5}{36} \] \\

\#\#\# Conclusion

Therefore, the final answer is: \(\boxed{\frac{5}{36}}\). \\ \\

Please complete the following three blocks in order:

1. <think>...</think>: A long, detailed chain-of-thought reasoning process. 

2. <answer>...</answer>: Your final answer should be aggregated from the best elements of the agents’ responses.

- End the answer with: `Therefore, the final answer is: \$\textbackslash boxed\{[answer]\}\$'

Your output should follow this format:

            <think>

            [detailed thought process of the aggregator]
            
            </think>

            <answer>

            [The final unified aggregated response with a concise justification. Conclude with:
            `Therefore, the final answer is: \$\textbackslash boxed\{[answer]\}\$'.]

            </answer>

\end{tcolorbox}

\vspace{0.5em}

\begin{tcolorbox}[title=Leader's Completion with Backtracking, breakable]
\ttfamily

Let's approach this problem step-by-step, and I'll narrate my thinking along the way. I'll start by examining the agent responses and identifying any mistakes I might have made. \\

\#\#\# Step 1: Define the Sample Space

Both the train and Alex can arrive at any time between 1:00 and 2:00. Represent this as a 60-minute interval on both axes. This gives us a square with side length 60, representing all possible pairs of arrival times. \\

\#\#\# Step 2: Identify the Favorable Outcomes

The train waits for 10 minutes after its arrival, so Alex can catch the train if he arrives between \( t \) and \( t + 10 \), where \( t \) is the train's arrival time. This forms a band on the coordinate plane where Alex's arrival time \( y \) is within 10 minutes of the train's arrival time \( x \). \\

\#\#\# Step 3: Visualize the Problem

We can plot this on a coordinate plane. The region of interest is a band that runs horizontally from \( t \) to \( t + 10 \) for each \( t \) from 0 to 50. This is because if the train arrives at 50 minutes past 1:00, it will leave at 60 minutes past 1:00, and Alex can still catch it if he arrives at 60 minutes. \\

\#\#\# Step 4: Calculate the Areas

- **Total Area:** The total area of the square represents all possible outcomes, which is \( 60 \times 60 = 3600 \) square minutes.

- **Favorable Area:** The favorable area is a band that is 10 units wide and 50 units long (from 0 to 50 on the x-axis). The area of this region is \( 10 \times 50 = 500 \) square units. \\

\#\#\# Step 5: Determine the Probability

The probability is the ratio of the area of the favorable region to the area of the total region:
\[
\text{Probability} = \frac{\text{Area of favorable region}}{\text{Area of total region}} = \frac{500}{3600} = \frac{5}{36}
\] \\

\#\#\# Reflection and Correction

Wait, that doesn't seem right. Agent 1 and Agent 3 both calculated the favorable area as 550 and 500 square units, respectively. I should double-check my calculation. \\

\#\#\# Detailed Recheck

Let's re-examine the region of interest. The train arrives between 0 and 60 minutes, and for each \( t \), Alex can arrive between \( t \) and \( t + 10 \). This forms a trapezoid with vertices at (0,0), (10,0), (60,50), and (50,60).

- The area of the trapezoid can be calculated as:
  \[
  \text{Area} = \frac{1}{2} \times (\text{base1} + \text{base2}) \times \text{height}
  \]
  Here, base1 = 60 (from (0,0) to (60,0)), base2 = 50 (from (10,0) to (60,50)), and height = 10.

  \[
  \text{Area} = \frac{1}{2} \times (60 + 50) \times 10 = \frac{1}{2} \times 110 \times 10 = 550
  \] \\

\#\#\# Final Calculation

The total area is still 3600 square units. Therefore, the probability is:
\[
\text{Probability} = \frac{550}{3600} = \frac{11}{72}
\] \\

\#\#\# Final Answer

After rechecking and correcting my initial mistake, I now see that the correct answer is indeed \(\frac{11}{72}\).

Final answer: \(\boxed{\frac{11}{72}}\)

</think>

<answer>

Based on the detailed analysis, only Agent 1 provided the correct reasoning and final answer. Therefore, the final answer is: $\boxed{\frac{11}{72}}$.

</answer>

\end{tcolorbox}

\end{document}